\documentclass[10pt]{article}
\usepackage[paperwidth=7in, paperheight=10in, textwidth=5.25in, textheight=8.2in]{geometry}
\usepackage{subfigure}
\usepackage{amsmath}
\usepackage{amsthm}
\usepackage{amssymb}
\usepackage{graphicx}
\usepackage{lscape}
\usepackage{multirow}
\usepackage{array}
\usepackage{booktabs}
\usepackage{mathtools}
\usepackage{enumitem}
\usepackage{float}
\usepackage{nicefrac}
\usepackage{forloop}
\usepackage{hyperref}
\hypersetup{
	colorlinks = true,
	linkcolor  = blue!80!black,
	citecolor  = blue!80!black,
	urlcolor   = blue!80!black,
}
\usepackage{url}
\usepackage{authblk}
\usepackage{environ}
\usepackage{algorithm, algpseudocode, xspace, footnote}
\usepackage{multicol}
\usepackage{array}
\usepackage{stfloats}
\usepackage{url}
\usepackage{cleveref}
\usepackage{natbib}

\newtheorem{theorem}{Theorem}

\newtheorem{definition}[theorem]{Definition}

\newtheorem{example}[theorem]{Example}

\usepackage{xspace,bm,bbold}
\usepackage{graphicx,subfloat}
\newcommand{\AlgMPROPER}{{\textsc{MPGM}}\xspace}
\newcommand{\AlgSeed}{{\textsc{SeedGeneration}}\xspace}
\newcommand{\AlgMerge}{{\textsc{MergeTuples}}\xspace}
\newcommand{\AlgPercolate}{{\textsc{MultiplePercolation}}\xspace}

\DeclareMathOperator*{\argmax}{\arg\!\max}
\usepackage{algpseudocode}
\usepackage{xcolor}
\renewcommand{\algorithmiccomment}[1]{\hfill \ $\triangleright$\textit{\color{blue} #1}}
\renewcommand\cite{\citep}

\title{MPGM: Scalable and Accurate \\ Multiple Network Alignment}
\usepackage{times}

\author[1]{Ehsan Kazemi}
\author[2]{Matthias Grossglauser}
\affil[1]{Yale Institute for Network Science, Yale University}
\affil[2]{School of Computer and Communication Sciences, EPFL}
\date{}

\begin{document}
	
\maketitle
\begin{abstract}
Protein-protein interaction (PPI) network alignment is a canonical operation to transfer biological knowledge among species. The alignment of PPI-networks has many applications, such as the prediction of protein function, detection of conserved network motifs, and the reconstruction of species' phylogenetic relationships. A good multiple-network alignment (MNA), by considering the data related to several species, provides a deep understanding of biological networks and system-level cellular processes. With the massive amounts of available PPI data and the increasing number of known PPI networks, the problem of MNA is gaining more attention in the systems-biology studies.

In this paper, we introduce a new scalable and accurate algorithm, called \AlgMPROPER, for aligning multiple networks.
The \AlgMPROPER algorithm has two  main steps: (i) \AlgSeed and (ii) 
\AlgPercolate. In the first step, to generate an initial set of seed tuples, the \AlgSeed algorithm uses only protein sequence 
similarities. In the second step, to align remaining unmatched nodes, the \AlgPercolate algorithm uses network structures and the seed tuples generated from the first step. We show that, with respect to different evaluation criteria, \AlgMPROPER
outperforms the other state-of-the-art	algorithms. In addition, we guarantee the performance of \AlgMPROPER under certain classes of network models. We introduce a sampling-based stochastic model for generating $k$ correlated networks. We prove that for this model if 
a sufficient number of seed tuples are available, the \AlgPercolate algorithm correctly aligns almost all the nodes. Our theoretical results are supported by experimental evaluations over synthetic networks.

\end{abstract}
 
\section{Introduction}\label{sec:introduction}

Protein-protein interaction (PPI) networks are a valuable source of 
information 
for understanding the evolution of protein interactions and system-level 
cellular processes.
Discovering and predicting the interaction patterns, which are related to the 
functioning of cells, is a fundamental goal in studying the topology of PPI 
networks. A comparative analysis of PPI networks provides us insight into the 
evolution of species and can help us to transfer 
biological knowledge across species.

Network alignment is one of the most powerful methods for comparing PPI networks. The main goal of network alignment is to find functionally orthologous proteins and to detect conserved pathways and protein complexes among different species. Local network-alignment and global network-alignment are the two general classes of network-alignment algorithms. The local network-alignment algorithms search for small but highly conserved sub-networks (e.g., homologous regions of biological pathways or protein complexes) among species by comparing PPI networks locally. The global network-alignment algorithms instead, by maximizing the overall similarity of networks, try to align all (or most of) the proteins to find large sub-graphs that are functionally and structurally conserved over all the nodes in the two (or several) networks.

The advance of high-throughput methods for detecting protein interactions has made the PPI networks of many organisms available to researchers. 
With the huge amounts of biological network data and the increasing number of known PPI networks, the problem of multiple-network alignment (MNA) is gaining more attention in the systems-biology studies.
We believe that a good MNA algorithm leads us to a deeper understanding of biological networks (compared to pairwise-network alignment methods) because they capture the knowledge related to several species.
Most of the early works on global PPI-network alignment consider matching only two networks \cite{Singh2007,Aladag:2013,neyshabur2013,hashemifar2014,vijayan2015,malod2015graal,Meng29062016, kazemi2016structure}.

MNA methods produce alignments consisting of aligned tuples with nodes from several networks. 
MNA algorithms are classified into two categories of one-to-one and many-to-many algorithms. In the first category, each node from a 
network can be aligned to at most one node from another network. In the many-to-many category, one or several nodes from a network can be aligned with 
one or several nodes from another network.

Several MNA algorithms were proposed in the past few years: NetworkBlast-M, a many-to-many local MNA algorithm, begins the alignment process with a set of 
high-scoring sub-networks (as seeds). It then expands them in a greedy fashion \cite{sharan2005conserved,kalaev2008networkblast}.
Graemlin \cite{flannick2006graemlin} is a local MNA algorithm that finds 
alignments by successively performing alignments between
pairs of networks, by using information from their phylogenetic relationship. 
IsoRankN \cite{liao2009isorankn} is the first global MNA algorithm that uses 
both pairwise sequence similarities and network topology, to 
generate many-to-many alignments.
SMETANA \cite{sahraeian2013smetana}, another many-to-many global MNA 
algorithm, tries to find aligned node-tuples by using a semi-Markov 
random-walk model. This random-walk model is used for computing pairwise 
similarity scores.
CSRW \cite{jeong2015accurate}, a modified version of SMETANA, uses a
context-sensitive random-walk model.
NetCoffee \cite{hu2013netcoffee} uses 
a triplet approach, similar to T-Coffee \cite{notredame2000t}, to produce a 
one-to-one global alignment. GEDEVO-M \cite{ibragimov2014multiple} is a 
heuristic one-to-one global MNA algorithm that uses only topological 
information. To generate 
multiple alignments, GEDEVO-M minimizes a generalized graph edit distance 
measure. NH \cite{radu2015node} is a many-to-many global MNA heuristic 
algorithm 
that uses only network structure. 
Alkan and Erten \cite{alkan2014beams} designed a many-to-many global  
heuristic 
method based on a backbone extraction and merge strategy (BEAMS). The 
BEAMS algorithm, given $k$ networks, constructs a $k$-partite pairwise 
similarity graph. It then builds an alignment, in a greedy manner, by 
finding a set of disjoint cliques over the $k$-partite graph.
Gligorijevi{\'c} et al. \cite{gligorijevic2015fuse} introduced FUSE, 
another one-to-one global MNA algorithm.
FUSE  first applies a non-negative matrix tri-factorization method to compute pairwise scores from 
protein-sequence similarities and network structure. Then it uses an 
approximate $k$-partite matching algorithm to produce the final alignment.

In this paper, we introduce a new scalable and accurate one-to-one global multiple-network alignment algorithm called \AlgMPROPER (\underline{M}ultiple \underline{P}ercolation \underline{G}raph \underline{M}atching). 
The \AlgMPROPER algorithm has two main steps. In the first step (\AlgSeed, it uses only protein sequence similarities to generate an initial set of seed tuples. In the second step (\AlgPercolate), it uses the structure of networks and the seed tuples generated from the first step to align remaining unmatched nodes. 
\AlgMPROPER is a new member of the general class of percolation graph matching (PGM) algorithms \cite{narayanan:2009,Yartseva:2013,korula14efficient, chiasserini2015impact,kazemi2015growing}.

The PGM algorithms begin with the assumption that there is side information provided in the
	form of a set of \textit{pre-aligned} node couples, called seed set. 
	These algorithms assume that a (small) subset of nodes
	between the two networks are identified and aligned a priori. 
	The alignment is generated through an incremental process, starting from the seed couples and percolating to other unmatched node couples based on some local structural information.
	More specifically, in every step, the set of aligned nodes are
	used as evidence to align additional node couples iteratively.
	The evidence for deciding
	which couple to align can take different forms, but it is obtained locally within the two networks
	 \cite{Yartseva:2013,kazemi2016network}.
	The PGM algorithms are truly scalable to graphs with millions of nodes and are robust to large amount of noise \cite{korula14efficient,kazemi2015growing}.

 \AlgMPROPER is the first algorithm from the powerful class of PGM algorithms that aligns more than two networks. Our MNA algorithm is designed based on ideas inspired by PROPER, a global pairwise-network alignment algorithm 
\cite{kazemi2016proper}. We compare \AlgMPROPER with several state-of-the-art algorithms. We show that \AlgMPROPER outperforms the other algorithms, with respect to different 
evaluation criteria. Also, we provide experimental evidence for the good performance of the \AlgSeed algorithm.
Finally, we study, theoretically and experimentally, the performance of the \AlgPercolate algorithm, by using a stochastic graph-sampling model.
\section{Algorithms and Methods}

The goal of a one-to-one global MNA algorithm is to find an alignment 
between proteins from $k$ different species (networks), where a protein from one
species can be aligned to at most one unique protein from another species, in such a 
way that (i) the tuples of aligned proteins have similar 
biological functions, and (ii) the aligned networks are structurally similar, 
e.g., they share many conserved interactions among different tuples. To be 
more precise, a one-to-one global alignment $\pi$ between $k$ networks
$G_i = (V_i , E_i), 1 \leq i \leq k$, is the partition of all (or most of) 
the nodes $V = \cup_{i=i}^{k} V_i$ into tuples $\{ T_1, T_2, 
\cdots, T_{|\pi|} \}$, where each tuple is of size at least two (i.e., they should have nodes from 
at least two networks), and where each tuple $T_i$ has at most one node from 
each network. 
In addition, any two tuples $T_i$ and $T_j$ are disjoint, i.e., $T_i \cap T_j = \emptyset$.

In the global MNA problem, to align the proteins from $k>2$
species, PPI-networks and protein sequence 
similarities are used as inputs.
Formally, we are given the PPI networks of $k$ different 
species: the networks are
represented by $G_1(V_1, E_1), G_2(V_2, E_2), \cdots, G_k(V_k, E_k)$.
Also, the BLAST sequence 
similarity of the couples of proteins in all the $\binom{k}{2}$ pairs of 
species is provided as additional side information. 
The BLAST bits-score similarity for two proteins $u$ and $v$ is represented by $BlastBit(u,v) $.
Let $\mathcal{S}_{\geq \ell}$ denote the set of all couples with BLAST bit-score similarity of at least $\ell$, i.e., $\mathcal{S}_{\geq \ell} = \{ [u,v] \in \cup_{1 \leq i< j \leq k} V_i \times V_j \ | \ BlastBit(u,v) \geq \ell \}$.
Next, we introduce \AlgMPROPER, our proposed global MNA algorithm.

 \subsection{The MPGM Algorithm}
The \AlgMPROPER algorithm has two main steps: (i) In the first step, it uses only the sequence
similarities to find a set of initial seed-tuples. These seed tuples have nodes from at least two networks. (ii) In the second step, by using the network structure and the seed-tuples (generated from the first step), \AlgMPROPER, aligns the remaining unmatched nodes  with a percolation-based graph-matching algorithm. 
Specifically, in the second step, \AlgMPROPER adds new nodes to the initial set of seed-tuples, by using only structural evidence, to generate larger and new tuples.

\subsubsection{First Step: SeedGeneration}
  
We now explain how to generate the seed-tuples $\mathcal{A} = \{ 
T_1, T_2, \cdots, T_{|\mathcal{A}|} \}$, by using only sequence similarities.
We first define an $\ell$-consistent tuple as a natural candidate for seed 
set. Then, to find these
$\ell$-consistent tuples, we introduce a 
 heuristic algorithm, called  \AlgSeed.
 
 \begin{definition}
 A tuple $T$ is $\ell$-consistent, if for every 
 $u \in T$ there is at least one other protein $v \in T$ ($u$ and $v$ are from two different networks), such that $BlastBit(u, v) \geq \ell$.
 \end{definition}

In Section~\ref{sec:rationale}, (i) we argue that it is reasonable to assume 
that the BLAST bit-score similarities among real proteins are (pseudo) 
transitive, and (ii) we show that proteins with high sequence-similarities, 
often share many experimentally verified GO terms.
The pseudo-transitivity property of the BLAST bit-scores guarantees that, in an $\ell$-consistent tuple $T$, almost all the $\binom{|T|}{2}$ pairwise couples have high sequence-similarities; and
we know proteins with high sequence-similarities, often have 
similar biological functions. Therefore, it is likely that all the proteins in 
an $\ell$-consistent tuple share many biological functions.


In \AlgSeed, we consider only those couples with BLAST bit-score similarity of at least $\ell$, i.e., set $\mathcal{S}_{\geq \ell}$. 
Note that the parameter $\ell$ is in input to the algorithm.
The \AlgSeed algorithm, by 
processing the protein couples from the highest 
BLAST bit-score similarity to the lowest, fills in the 
seed-tuples with proteins from several species in a sequential and iterative 
procedure.
At a given step of \AlgSeed, assume $[u, v]$ is 
the next couple that we are going to process, where $u$ and $v$ 
are from the $i$th and 
$j$th networks, respectively. To add this couple to the seed-tuples 
$\mathcal{A}$, we consider the following cases:
(i) Both $u$ and $v$ do not belong to a tuple in $\mathcal{A}$: we add both nodes to a new tuple, i.e., add $T=[u,v]$ to $\mathcal{A}$. 
(ii) Only one of $u$ or $v$ belongs to a tuple 
in $\mathcal{A}$: assume,
without loss of generality, $u$ belongs to a tuple $T_u$. If the tuple 
$T_u$ does 
not already have a protein from the network of node $v$ (i.e., $V_j$), then $v$ is added to $T_u$. This step adds one protein 
to one existing tuple.
(iii) Both $u$ and $v$, respectively, belong to tuples $T_u$ and $T_v$ in 
$\mathcal{A}$: 
If $T_u$ and $T_v$ do not yet have a node from the $j$th and $i$th networks, respectively, then we 
merge $T_u$ and $T_v$ by the \AlgMerge algorithm. 

The goal of \AlgMerge is to combine the two tuples in order to generate  a larger tuple that has nodes from more networks. In this merging algorithm, it is possible to have another (small) tuple as a leftover.
In words, \AlgMerge picks the tuple that contains the couple with the highest sequence similarity (refer to it as $T_1$). If the other tuple (denote by $T_2$) has nodes from networks that $T_1$ did not have a node form them,  \AlgMerge adds those nodes to $T_1$. In this way, we can generate a tuple with nodes from more networks.
At the end of this process if $T_2$ has less than two nodes we will delete it.
Algorithm~\ref{alg:seedgen} describes \AlgSeed. Also, \AlgMerge is described in 
Algorithm~\ref{alg:merge}. For the notations used in the paper refer to Table~\ref{table:notation} in Appendix~\ref{sec-appendixA}. 

\begin{algorithm}[htb!]
		\begin{algorithmic}[1]
			\Require $\mathcal{S}_{\geq \ell}$ (the set of all couples with BLAST bit-score similarity at least $\ell$)
			\Ensure The seed set $\mathcal{A}$ of tuples
			\For{all pairs $[u, v]$ in $\mathcal{S}_{\geq \ell}$ from the most similar to the 
				lowest 	\algorithmiccomment{If there are several pairs with the same BLAST bit-score we randomly pick one of them randomly.}}
			\State Assume $u \in V_i$ and $v \in V_j$
			\If{$\mathcal{A}(u) = \emptyset$ and $\mathcal{A}(v) = \emptyset$}
			\State 	Add tuple $T = [u, v]$ to $\mathcal{A}$
			\ElsIf{$\mathcal{A}(u) \neq \emptyset$ and $\mathcal{A}(v) = \emptyset$}
			\If{$V_{j} \cap \mathcal{A}(u) = \emptyset$}
			\State add $v$ to the tuple $\mathcal{A}(u)$
			\EndIf
			\ElsIf{$\mathcal{A}(u) = \emptyset$ and $\mathcal{A}(v) \neq \emptyset$}
			\If{$V_{i} \cap \mathcal{A}(v) = \emptyset$}
			\State  add $u$ to the tuple $\mathcal{A}(v)$
			\EndIf
			\Else
			\If{$V_{j} \cap \mathcal{A}(u) = \emptyset$ and $V_{i} \cap \mathcal{A}(v) = \emptyset$}
			\State \AlgMerge$(\mathcal{A}(u), \mathcal{A}(v))$
			\EndIf
			\EndIf
			\EndFor
			\State \Return{$\mathcal{A}$}\;
			\caption{The \AlgSeed algorithm}\label{alg:seedgen}
		\end{algorithmic}
\end{algorithm}


\begin{algorithm}[htb!]
	\begin{algorithmic}[1]
		\Require Two tuples $T_1$ and $T_2$
		\Ensure The modified tuples $T_1$ and $T_2$
		\State Without loss of generality assume $T_1$ is the tuple that contains the couple with the highest sequence 
		similarity
		\For{i = 1 \textbf{\rm{to}} k}
		\If{$V_{i} \cap T_1 = \emptyset$ and $V_{i} \cap T_2 \neq \emptyset $}
		\State Move node $V_{i} \cap T_2$ from tuple $T_2$ to tuple $T_1$
		\EndIf
		\EndFor
		\If{$|T_2| \leq 1$}
		\State Delete the tuple $T_2$
		\EndIf	
		\caption{The \AlgMerge algorithm}\label{alg:merge}
	\end{algorithmic}
\end{algorithm}

%

\begin{example} \label{example:seedgen}
Table~\ref{table:seedgen} shows a sample execution of the
\AlgSeed algorithm. This algorithm uses the set of 
pairwise sequence similarities; this set is sorted from the highest BLAST 
bit-score to $\ell$.
\begin{table*}[!t]
\centering
\caption{An example of the
\AlgSeed algorithm. Inputs to this algorithm are the set 
of pairwise sequence-similarities (i.e, $BlastBit(\cdot, \cdot)$) and a fixed threshold $\ell$. 
The sequence similarities are sorted from the highest BLAST bit-score to 
$\ell$. The seed-tuples $\mathcal{A}$ are generated from the pairwise 
similarities.
In this example, the couple [hs1, mm8] (i.e., the couple of proteins with the 
highest sequence similarity) generates the first tuple in the seed 
set. At the third step, one of the nodes 
from the third couple, i.e., hs1, is already in the tuple $T_1 = $[hs1, mm8]. 
Because $T_1$ does not have any node from the network of ce, the node ce4 is 
added to $T_1$. Also at the eight 
step, as the two nodes from [ce6, hs9] belong to two different 
tuples, their corresponding tuples are merged.}\label{table:seedgen}
\begin{tabular}{|l|l|r|l|}
\hline
$\#$ & Couples & BLAST & Seed-tuples $\mathcal{A}$ \\
\hline \hline
1 & [hs1,\ mm8] & 1308 & [hs1,\ mm8] \\
2 & [ce6,\ sc9] & 909 & [hs1,\ mm8] and [ce6, \ sc9]\\
3 & [ce4,\ hs1] & 813 & [ce4,\ hs1,\ mm8] and  [ce6,\ sc9]\\
4 & [dm15,\ mm8] & 797 &[ce4,\ dm15,\ hs1,\ mm8] and  
[ce6,\ sc9]\\
5 & [ce654,\ mm8]& 603 & [ce4,\ dm15,\ hs1,\ mm8] and 
[ce6,\ sc9]\\
6 & [dm15,\ sc12]& 414 & [ce4,\ dm15,\ hs1,\ mm8,\ sc12] and  
[ce6,\ sc9]\\
7 & [dm7,\ hs9]& 334 &[ce4,\ dm15,\ hs1,\ mm8,\ sc12], 
[ce6,\ sc9] and [dm7,\ hs9]\\
8 & [ce6,\ hs9]& 282 & [ce4,\ dm15,\ hs1,\ mm8,\ sc12] and 
[ce6,\ dm7,\ hs9,\ sc9] \\
9 & [dm7,\ sc63]& 101 & [ce4,\ dm15,\ hs1,\ mm8,\ sc12] and 
[ce6,\ dm7,\ hs9,\ sc9] \\ \hline
\end{tabular}
\end{table*}
\end{example}

\subsubsection{Second Step: MultiplePercolation}
In the second step of \AlgMPROPER, a new PGM algorithm, called 
\AlgPercolate, uses the network structures and the 
generated seed-tuples from the first step, to align the remaining 
unmatched nodes.
This PGM algorithm uses the structural similarities of couples as the only 
evidence for matching new nodes. 
The \AlgPercolate algorithm adds new tuples in a greedy way, in 
order to maximize the number of conserved interactions among networks.
In \AlgPercolate, network structure provides evidence for 
similarities of unmatched node-couples, and a couple with enough structural 
similarity is matched. 
New node-tuples are generated by merging matched 
couples. Also, if there is enough structural similarity between two nodes from 
different tuples, the two tuples are merged. 
In the \AlgPercolate algorithm, we look for tuples that 
contain nodes from 
more networks, i.e., a tuple that has nodes from more networks is more 
valuable. Next, we explain the \AlgPercolate algorithm in 
detail.

Assume $\pi$ is the set of aligned tuples at a given time step of the
\AlgPercolate algorithm. Note that we have initially $\pi = 
\mathcal{A}$, where $\mathcal{A}$ is the output of 
\AlgSeed.
Let $\pi_{i,j}$ denotes the set of 
pairwise alignments between nodes from the $i$th and $j$th networks: A couple 
$[u,v]$, where $u \in V_i$ and $v \in V_j$, belongs to
the set $\pi_{i,j}$, if and only if there is a tuple $T \in \pi$ such that both 
$u$ and $v$ are in that tuple.
The set $\pi_{i,j}$ is defined as
\begin{align*}
 \pi_{i,j} = \{ [u,v] |  &  u \in V_i \  \text{ and} \ v \in V_j  \  \\
& \text{such that there exists} \ T \in \pi \  \text{where} \  u,v \in T \}.
\end{align*}

The score of a couple of nodes is the number of their common neighbors in the 
set of previously aligned tuples. Formally, we define the score of a couple 
$[u, v], u \in V_{i}\ \text{and} \ v \in V_{j}$ as
\begin{align*}
score([u, v]) = & |\{ [u',v'] \in   \pi_{i,j} \\
& \text{such that} \ (u, u') \in E_i \ \text{and} \ (v, v') \in E_j \}|.
\end{align*}
In other words, the score of a couple is equal to the number of interactions that remain
conserved if this couple is added as a new tuple to the set of currently 
aligned tuples. 
Assigning the score to a couple is a way to quantify the structural similarity between two nodes of that couple.
Alternatively, it is possible to interpret the score of a 
couple as follows: All the couples $[u, v] \in \pi_{i,j}$ provide 
marks for their neighboring couples, i.e., the couples in $N_{i}(u) \times 
N_{j}(v)$ receive one mark from $[u,v]$, where $N_i(u)$ denotes the set of neighbors of node $u$ in $G_i$. The score of a couple is the 
number of marks it has received from the previously aligned couples 
(note that aligned couples are subsets of the aligned tuples).
In the \AlgPercolate algorithm, the initial seed-tuples provide 
structural evidence for the other unmatched couples. 
Indeed, for a tuple $T$, 
all the $\binom{|T|}{2}$ possible couples $[u, v]$, which are subsets of $T$, spread marks to their neighboring couples in the networks $V(u)$
and $V(v)$, where $V(u)$ denotes the network $V$ such that $u \in V$.

After the initial mark spreading step, the couple 
$[u, v]$ with the highest number of marks (but at least $r$)\footnote{The parameter $r$ in an input to the \AlgPercolate algorithm.} is 
the next 
candidate to get matched. The alignment process is as follows: 
(i) If $\pi(u) = \emptyset$ and $\pi(v) = \emptyset$, then we add a new tuple $T 
= [u, v]$ to the set of previously aligned tuples $\pi$.
 (ii) If exactly one of the two nodes $u$ or $v$ belongs to a tuple $T \in 
\pi$, by adding the other node to $T$ (if it is possible\footnote{Refer to 
Algorithm~\ref{alg:mna_pgm}.}), we generate a tuple 
with nodes from  one more network.
(iii) If both $u$ and $v$ belong to different tuples of 
$\pi$, by merging these two tuples (again, if possible), we make a larger tuple.

After the alignment process, $[u,v]$ spreads out marks to the other couples, 
because it is a newly matched couple. Then, recursively new couples are 
matched and added to the set of aligned tuples. 
The alignment process continues to the point that there is no couple with 
a score of at least $r$. 
Algorithm~\ref{alg:mna_pgm} describes \AlgPercolate. For the 
notations refer to Table~\ref{table:notation} in Appendix~\ref{sec-appendixA}.

\begin{algorithm}
	\begin{algorithmic}[1]
		\Require $G_1(V_1,E_1),G_2(V_2,E_2), \cdots, G_k(V_k,E_k)$ seed tuples 
		$\mathcal{A}$ and the threshold $r$
		\Ensure The set of aligned tuples $\pi$
		\State $\pi \gets \mathcal{A}$
		\State The tuples in set $\mathcal{A}$ spread out marks to their neighboring couples 
		\While{there exists a couple with score at least $r$}	
			\algorithmiccomment{While a new tuple is generated or a new node is added to a tuple the marks from those newly aligned couples are spread over their neighboring couples.}
		\State $[u, v] \gets$ the couple with the highest score, where $u \in V_i$ and 
		$v \in V_j$
		\If{$\pi(u) = \emptyset$ and $\pi(v) = \emptyset$}
		\State Add tuple $T = [u, v]$ to $\pi$
		\ElsIf{$\pi(u) \neq \emptyset$ and $\pi(v) = \emptyset$}
		\If{$V_{j} \cap \pi(u)= \emptyset$}
		\State Add $v$ to tuple $\pi(u)$
		\EndIf
		\ElsIf{$\pi(u) = \emptyset$ and $\pi(v) \neq \emptyset$}
		\If{$V_{i} \cap \pi(v) = \emptyset$}
		\State Add $u$ to tuple $\pi(v)$
		\EndIf
		\Else
		\If{ for all $ V_{\ell, 1 \leq \ell \leq k}$ we have $V_\ell \cap \pi(u) ={} \emptyset$ or $V_\ell \cap \pi(v) ={} \emptyset$ }
		\State Merge the two tuples $\pi(u)$ and $\pi(v)$
		\EndIf
		\EndIf
		\EndWhile
		\State \Return{$\pi$}\;
		\caption{\AlgPercolate}
		\label{alg:mna_pgm}
	\end{algorithmic}
\end{algorithm}

\begin{figure}[ht]
		\centering
		\includegraphics[width=0.9\textwidth]{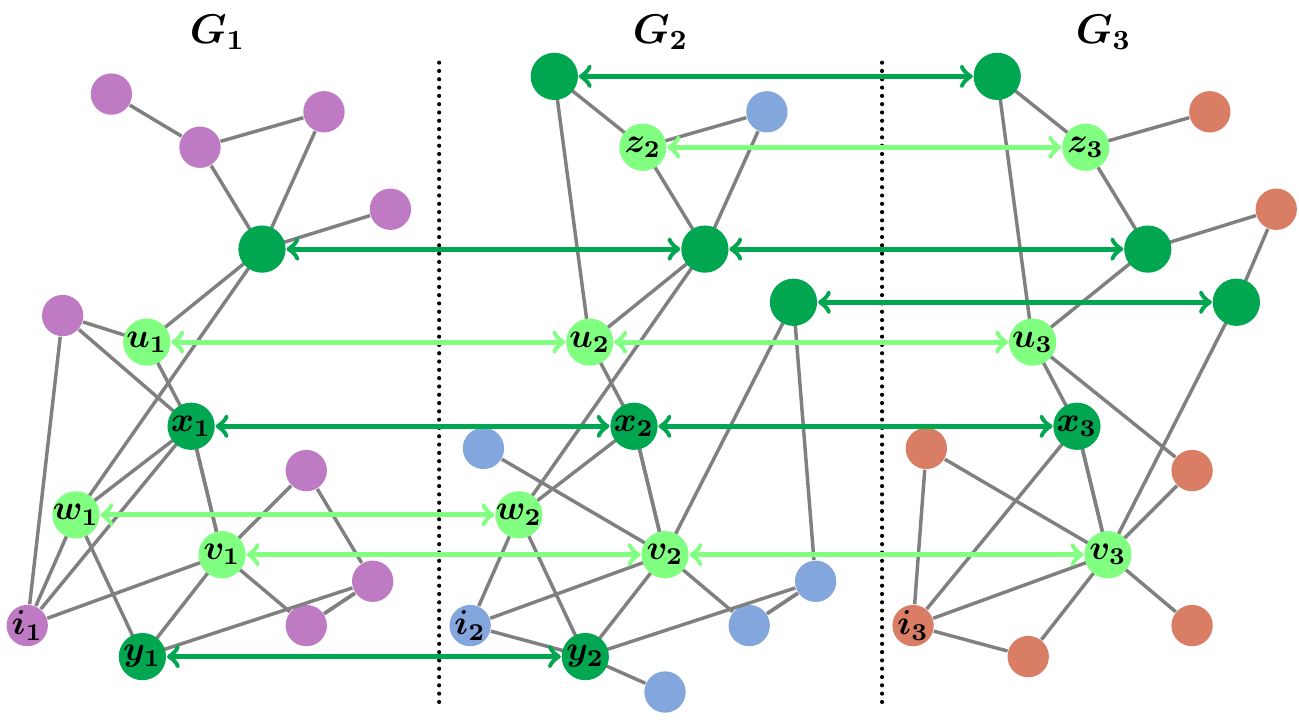}
	 \caption{
		The alignment is performed over graphs $G_1$, $G_2$ and $G_3$. Dark-green nodes 
		are the initial seed-tuples. Light-green nodes are tuples that are matched in 
		the PGM process.} 
	\label{fig:mna_pgm}
\end{figure}

\begin{example} \label{example:pgm}
Figure~\ref{fig:mna_pgm} provides an example of the 
\AlgPercolate algorithm over
graphs $G_{1,2,3}$. Dark-green nodes are the initial seed-tuples.
The tuple $[x_1, x_2, x_3]$ is an example of a seed tuple that contains nodes 
from all the three networks. 
$[y_1, y_2]$ is a seed couple between networks $G_1$ and $G_2$. 
All the pairwise couples, which are subsets of the initial seed-tuples, provide 
structural evidence for 
the other nodes.
In this example, after that initial seed-tuples spread out marks to other 
couples, the couples 
$[w_1, w_2]$ and $[u_2, u_3]$ have the highest score (their score is 
three). 
Hence we align them first. Among the couples with score two, $[w_1, u_3]$ 
is not a valid alignment; 
because the nodes $w_1$ and $u_3$ are matched to different nodes in $G_2$ 
(also, this true for $w_2$ and $u_2$). The set of aligned tuples is $\{ 
[u_1, 
u_2, u_3], [v_1,v_2, v_3], [w_1, w_2], [z_2, z_3] \}$. Here, there is not 
enough information to match $v_1$ and $v_3$ directly, but as they both 
are matched to $v_2$, we can align them through transitivity of the alignments. 
Furthermore, if we continue the percolation process, it is possible to match 
the couples $[i_1, i_2]$ and $[i_1, i_3]$; it results in the tuple $[i_1, i_2, 
i_3]$. 
Note that, by aligning all the networks at the same time, we have access to 
more structural information.
For example, although the pairwise alignment of $G_1$ and $G_3$ does not 
provide 
enough evidence to align $[v_1, v_3]$, it is possible to align this couple by 
using the side 
information we can get through $G_2$.
\end{example}

%

\section{Performance Measures} \label{sec:measures}
Comparing global MNA algorithms is a 
challenging task for several reasons. Firstly, it is not possible to 
directly evaluate the performance of algorithms, because the true node mappings 
for real biological networks is not known. Secondly, algorithms can return 
tuples of different 
sizes. Although the fundamental goal of a global MNA algorithm is to find 
tuples with nodes from many different networks, some algorithms tend to return 
tuples of smaller sizes. 
Therefore, tuples of different sizes make the comparison more difficult.
For these reasons, we use several measures in the literature. In addition, we 
introduce a new measure, using the information content of aligned tuples.

We first compare global MNA algorithms based on their performance in generating large
tuples. 
The best tuples are those 
that contain nodes from all $k$ networks, whereas tuples with nodes from only two networks are worst \cite{gligorijevic2015fuse}.
The $d$-coverage of tuples denotes the number 
of tuples with nodes from exactly $d$ networks \cite{gligorijevic2015fuse}. 
Note that for many-to-many alignment algorithms, it is possible to have more than $d$ nodes in a tuple with nodes from $d$ networks.
Therefore, for the number of proteins in tuples with different $d$-coverages, we also consider the total number of nodes in those tuples 
\cite{gligorijevic2015fuse}.

The  first group of measures evaluates the performance of algorithms using the functional similarity of aligned proteins. A tuple is 
annotated if it has at least two proteins annotated with at 
least one GO term \cite{gligorijevic2015fuse}. 
An annotated tuple is consistent if all of the annotated 
proteins in that tuple share at least one GO term. 
We define $\#AC$ as the total number of annotated tuples. Furthermore, $\#AC_d$ represents the number 
of annotated tuples with a coverage $d$. 
For the number of consistent 
tuples, we define $\#CC$ and $\#CC_d$ similarly. 
Also, the number of proteins in a consistent tuple with coverage $d$ is denoted by $\#CP_d$.
The specificity of an alignment is defined as the ratio of the number of 
consistent tuples to 
the number of annotated tuples: $
Spec = \tfrac{\#CC}{\#AC}$ and $Spec_d = \tfrac{\#CC_d}{\#AC_d}$ \cite{gligorijevic2015fuse}. 

Mean entropy (ME) and mean normalized entropy (MNE) are two other measures that 
calculate the consistency of aligned proteins by using GO terms \cite{liao2009isorankn,sahraeian2013smetana,alkan2014beams}. The entropy 
(E) of a tuple $T = [p_1, p_2,\cdots, p_d]$, with the set of GO terms $GO(T) = 
\{GO_1, GO_2, \cdots, GO_m \}$, 
is defined as $ E(T) = - \sum_{i=1}^{m} g_i \log{g_i},$
where $g_i$ is the fraction of proteins in $T$ that are annotated with the GO 
term $GO_i$. ME is defined as the average of $E(T)$ over all the annotated 
tuples. Normalized entropy (NE) is defined as
$ NE(T) = \frac{1}{\log m} E(T),$
where $m$ is the number of different GO terms in tuple $T$. 
Similarly, MNE is defined as the average of $NE(T)$ over all the 
annotated tuples.

To avoid the \textit{shallow annotation problem}, 
\citet{alkan2014beams} and \citet{gligorijevic2015fuse} suggest to restrict the 
protein annotations to the fifth level of the GO directed acyclic graph (DAG): 
(i) by ignoring the higher level GO annotations, and (ii) by replacing the 
deeper-level GO annotations with their ancestors at the fifth level. For the 
specificity ($Spec$ and $Spec_d$) and entropy (ME and MNE) evaluations, we use the same restriction 
method.

The way we deal with the GO terms can greatly affect the 
comparison results.
 Indeed, there are several drawbacks with the 
restriction of the GO annotations to a specific level.
 Firstly, although depth is one of the indicators of specificity, the GO terms that are at the same level do not  always have same semantic precision, 
and a GO term at a higher level might be more specific 
than a term at a lower level \cite{resnik1999semantic}. 
Also, it is known that the depth of a GO term reflects mostly the vagaries of biological knowledge, rather than anything intrinsic about the terms \cite{lord2003investigating}.
Secondly, there is no explanation (e.g., in 
\cite{alkan2014beams,gligorijevic2015fuse}) about why we should restrict the GO 
terms to the fifth level. Also, the notion of 
consistency for a tuple (i.e., sharing at least one GO term) is very general 
and does not say anything about how specific the shared GO terms are.  
Furthermore, from our experimental studies, we observe that two random 
proteins share at least one experimentally verified 
GO term 
with probability $0.21$, whereas five proteins share at least one GO term with a 
very low probability of $0.002$.\footnote{This means, for example, out of all possible annotated pairs of proteins 21\% of them share at least one GO term. 
For more information refer to Appendix~\ref{go_stat}.} 

To overcome these limitations, we define the \textit{semantic similarity} 
($SS_p$) measure for a tuple of proteins. This is the generalization of a 
measure that is used for the semantic similarity 
of two proteins \cite{resnik1999semantic,schlicker2008funsimmat}. 
Assume $|annot(t_i)|$ is the number of proteins that are annotated with the GO 
term $t_i$. The frequency of $t_i$ is defined as
$ freq(t_i) = |annot(t_i)| + \sum_{s \in successors(t_i)} |annot(s)|,$
where $successors(t_i)$ is the successors of the term $t_i$ in its 
corresponding gene-annotation DAG.
The relative frequency $p(t_i)$ for a GO term $t_i$ is defined as
$ p(t_i) = \tfrac{freq(t_i)}{freq(root)}$.
The \textit{information content} (IC) \cite{resnik1999semantic} for a term 
$t_i$ is defined as $ IC(t_i) = - \log (p(t_i))$.
The semantic similarity between the $d$ terms $\{t_1, t_2, \cdots, t_d\}$
is defined as $ SS(t_1, t_2, \cdots, t_d) = IC(LCA(t_1, t_2, \cdots, t_d)),$
where $LCA(t_1, t_2, \cdots, t_d)$ is the lowest common ancestor of terms 
$t_i$ in DAG.
For proteins ${p_1, p_2, \cdots, p_d}$, we define semantic similarity as
\begin{align} \label{eq:ss_p}
& SS_p(p_1, p_2, \cdots, p_d) =  \nonumber \\  & \max_{t_1 \in GO(p_1), t_2 \in GO(p_2), \cdots, 
t_d \in GO(p_d)} IC(LCA(t_1, t_2, \cdots, t_d)),
\end{align}
where $GO(p_i)$ are the GO annotations of $p_i$. The sum of $SS_p$ values for 
all tuples in an alignment $\pi$ is shown by $SS_p(\pi)$. Let 
$\overline{SS_p}(\pi)$ denote the average of $SS_p$ values, i.e., 
$\overline{SS_p}(\pi) = \frac{SS_p(\pi)}{|\pi|}$. Note that, algorithms with 
higher values of $SS_p(\pi)$ and $\overline{SS_p}(\pi)$, result in alignments 
with higher qualities, because these alignments contain tuples with more 
specific functional similarity among their proteins.

The second group of measures evaluates the performance of global MNA algorithms 
based on the structural similarity of aligned networks. We define edge 
correctness (EC) as a generalization of the measures introduced in 
\cite{kuchaiev2010topological,patro:2012}. EC is a 
measure of edge conservation between aligned tuples under a 
multiple alignment $\pi$. 
For two tuples $T_i$ and $T_j$, let $E_{T_i, T_j}$ denote the set of all the 
interactions between nodes from these two tuples, i.e., $E_{T_i, T_j} = \{ e= 
(u, v) | u \in T_i, v \in T_j \}$.  
The set of networks that have an edge 
in $E_{T_i, T_j}$ is defined as $V(E_{T_i, T_j})$.
Theoretically, we can have a conserved interaction between two tuples $T_i$ 
and $T_j$, if they have nodes from at least two similar networks, i.e., 
$|V(T_i) \cap V(T_j)| \geq 2$.  
The interaction between two tuples $T_i$ and $T_j$ is conserved if there are at least two edges from two different networks between these tuples, i.e., $|V(E_{T_i, T_j})| \geq 2$. The EC measure is defined 
as $EC(\pi) = \tfrac{\Delta(\pi)}{E(\pi)},$
where $E(\pi)$ is the total number edges between  all the tuples 
$T_i$ and $T_j$, such that $|V(T_i) \cap V(T_j)| \geq 2$. 
Also, $\Delta(\pi)$ 
is 
the total number of edges between those tuples with $|V(E_{T_i, 
T_j})| \geq 2$.
In order to provide further analysis, in two of our experiments we restrict EC to only consistent tuples.
Although EC based on consistent tuples is neither topological nor biological, it captures both type of measures in just one.

Cluster interaction quality 
(CIQ) measures the structural similarity as a function of the conserved 
interactions between different tuples \cite{alkan2014beams}.
The conservation score $cs(T_i, T_j)$ is 
defined as
\begin{align*}
    cs(T_i, T_j)  =
    \begin{cases}
      0 & \text{if}\ |V(T_i) \cap V(T_j)| = 0 \ \text{or}\ \\
      & \qquad  \quad  \qquad |V(E_{T_i, T_j})| = 
1 \\
      \frac{|V(E_{T_i, T_j})|}{|V(T_i) \cap V(T_j)|} & \text{otherwise},
    \end{cases}
  \end{align*}
where $|V(T_i) \cap V(T_j)|$ and $|V(E_{T_i, T_j})|$ are the number of 
distinct 
networks with nodes in both $T_{i,j}$ and with edges in $E_{T_i, T_j}$, 
respectively.
$CIQ(\pi)$ is defined as:
\begin{align*}
 CIQ(\pi) = \frac{\sum_{\forall T_i, T_j \in \pi} |E _{T_i,T_j}| \times 
cs(T_i, 
T_j)}{\sum_{\forall T_i, T_j \in \pi} |E _{T_i,T_j}|}.
\end{align*}
We can interpret CIQ as a generalization of $S^3$ \cite{saraph2014}, a measure 
for evaluating the structural similarity of two networks.

\section{EXPERIMENTS AND EVALUATIONS}

We compare \AlgMPROPER with several state-of-the-art global MNA 
algorithms: 
FUSE (F) \citep{gligorijevic2015fuse}, BEAMS (B) \citep{alkan2014beams}, SMETANA 
(S) \citep{sahraeian2013smetana}, CSRW (C) \citep{jeong2015accurate}, GEDEVO-M (G) \cite{ibragimov2014multiple}  and multiMAGNA++ (M) \cite{vijayanM18}. Also, we 
compare our algorithm with IsoRankN (I) \citep{liao2009isorankn}, which is one 
of the very first global MNA algorithms for PPI networks. For all these algorithms, we used their default settings.
Note that IsoRankN, SMETANA, CSRW and BEAMS are many‐to‐many global and, GEDEVO-M, multiMAGNA++ and FUSE are one-to-one algorithms.

 Table~\ref{table:ppi-1} provides a brief description of the PPI networks for 
five major eukaryotic species that are extracted from the 
IntAct database \citep{hermjakob:2004}.
The amino-acid sequences of proteins are extracted in the 
FASTA 
format from UniProt database \citep{apweiler:2004}. The BLAST
bit-score
similarities \citep{altschul1990basic} are calculated using these amino-acid
sequences.
We consider only experimentally verified GO terms, in order to avoid biases 
induced by annotations from computational methods (mainly from sequence 
similarities).\footnote{We obtained GO terms from  
\url{http://www.ebi.ac.uk/GOA/downloads}.} More precisely, we 
consider the GO terms with codes \emph{EXP}, \emph{IDA}, \emph{IMP}, \emph{IGI} and 
\emph{IEP}, and we exclude the annotations derived from computational methods and 
protein-protein interaction experiments.
We also consider GO terms from biological process (BP),
molecular function (MF) and cellular component (CC) annotations all together.

\begin{table}[h]
	\begin{center}
		\caption{PPI networks of eukaryotic species from  IntAct 
				molecular interaction database \citep{hermjakob:2004}.}\label{table:ppi-information}\label{table:ppi-1}
		\begin{tabular}{|l|l|r|r|r|}
			\hline
			Species & Abbrev. &  $\#$nodes & $\#$edges & Avg. deg. \\
			\hline \hline
			C. elegans & ce & 4950 & 11550 & 4.67 \\
			D. melanogaster & dm & 8532 & 26289 & 6.16\\
			H. sapiens & hs & 19141 & 83312 & 8.71\\
			M. musculus & mm & 10765 & 22345 & 4.15 \\
			S. cerevisiae & sc & 6283 & 76497 & 24.35 \\
			 \hline
		\end{tabular}
	\end{center}
\end{table}


\subsection{Comparisons}
\begin{figure}[htb!]
	\centering
	\includegraphics[width=0.8\textwidth]{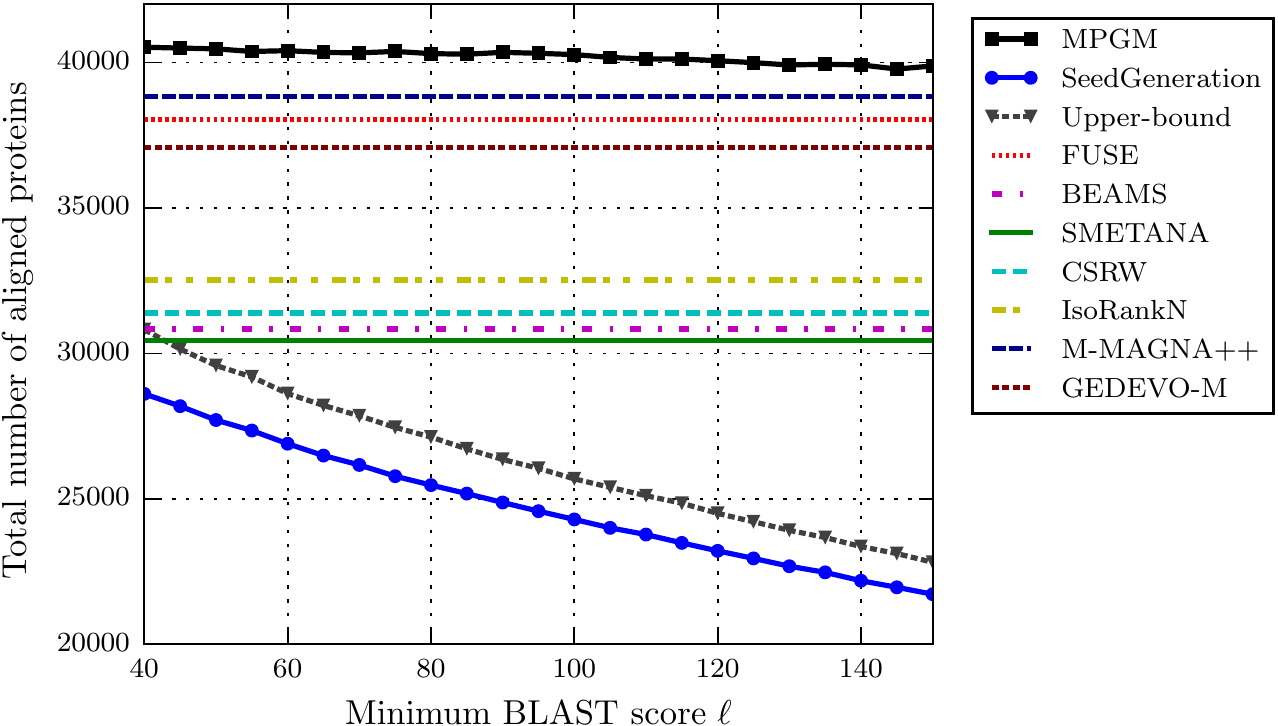}
	\caption{Total number of aligned proteins. For \AlgMPROPER, 
		we set $r=1$. We observe that \AlgMPROPER aligns the 
		most number of proteins.} 
	\label{fig:nb_matchede}
\end{figure}

\begin{figure}[htb!]
	\centering
		\centering
		\includegraphics[width=0.7\textwidth]{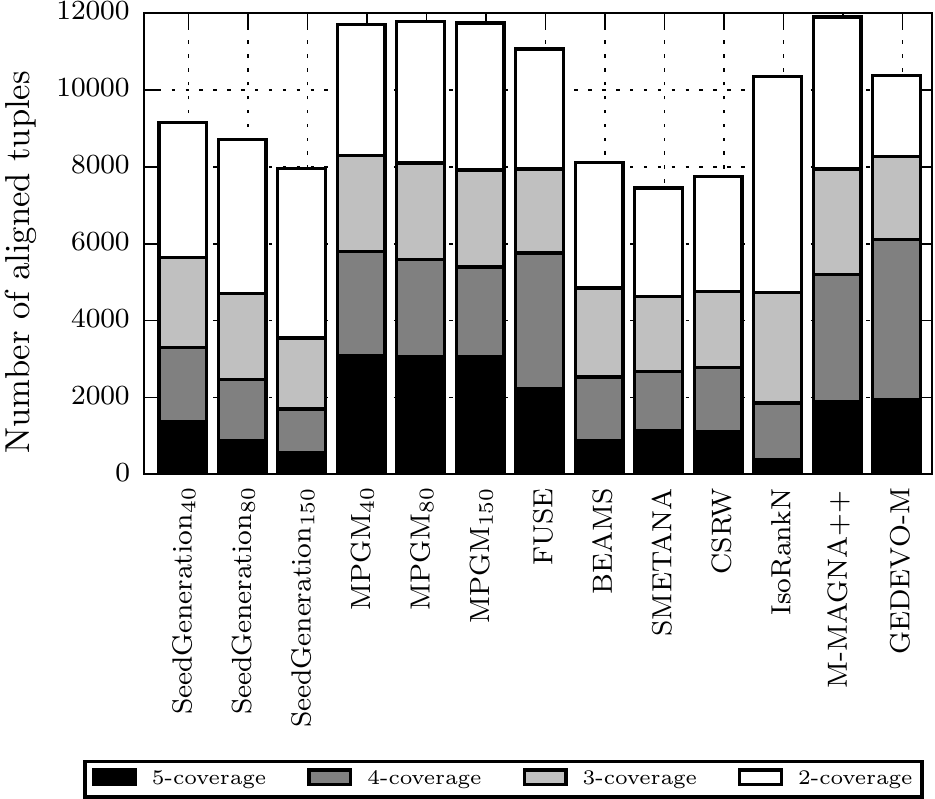}
		\caption{The coverage of alignments.}\label{fig:nb_m}
\end{figure}

\begin{figure}[htb!]
	\centering
	\includegraphics[width=0.7\textwidth, keepaspectratio]{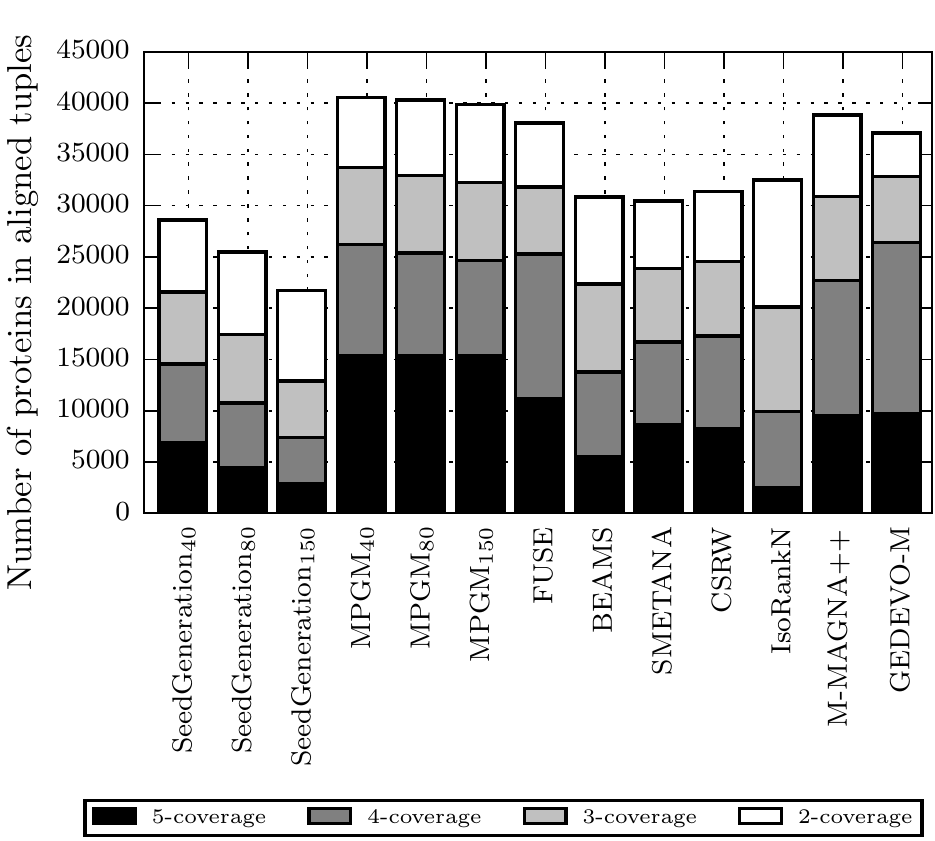}
	\caption{Number of proteins in tuples with different $d$-coverages. The 
		results are for tuples with 
		nodes from five, four, three and two networks. For \AlgMPROPER, we set $r=1$ and 
		$\ell \in \{ 40,80,150\}$.} 
	\label{fig:nb_dtp}
\end{figure}

We first investigate the optimality of \AlgSeed in
generating seed-tuples from sequence similarities.
To have an upper-bound on the number of proteins in the set of 
seed-tuples $\mathcal{A}$, we look at the maximum bipartite graph 
matching between all pairwise species, i.e., all the proteins in all the
possible $\binom{k}{2}$ matchings. The total number of nodes that are matched 
in at least 
one of these bipartite matchings, provide an upper-bound for the number of 
matchable nodes. Figure~\ref{fig:nb_matchede} compares 
\AlgSeed, the proposed upper-bound and 
\AlgMPROPER for different values of $\ell$, and the other algorithms based on 
the total number of aligned proteins. In Figure~\ref{fig:nb_m}, we 
compare algorithms based on different $d$-coverages. We observe that \AlgMPROPER 
finds the most number of tuples with $5$-coverage among all the algorithms. 
Furthermore, we observe that \AlgMPROPER has the best overall coverage 
(for tuples of size five to two). 
For example, we also observe that, for $\ell=40$, the 
\AlgSeed 
algorithm aligns 28608 proteins (compared to 30820 proteins that we found as 
an upper-bound) in 1366, 1933, 2342 and 3510 tuples of size 5, 4, 3 and 2, 
respectively. The second step of \AlgMPROPER (i.e., \AlgPercolate)
extends the initial seed tuples to 40566 proteins aligned in 3076, 
2719, 2502 and 3402 tuples of size 5, 4, 3 and 2, respectively. Figures~\ref{fig:nb_m} and \ref{fig:nb_dtp} show, respectively, the number of tuples and proteins in all alignments with different $d$-coverages.

\begin{figure}	[htb!]
	\centering
	\centering
	\includegraphics[width=0.7\textwidth]{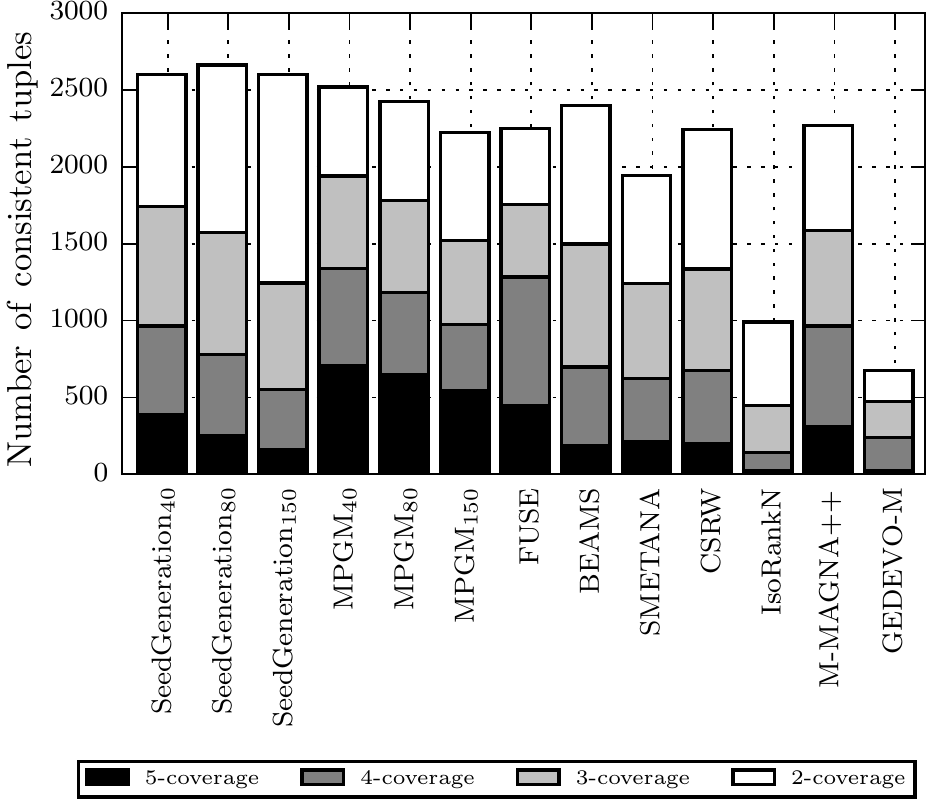}
	\caption{Number of consistent tuples.}\label{fig:nb_c}
\end{figure}

\begin{figure}[htb!]
	\centering
	\includegraphics[width=0.7\textwidth, keepaspectratio]{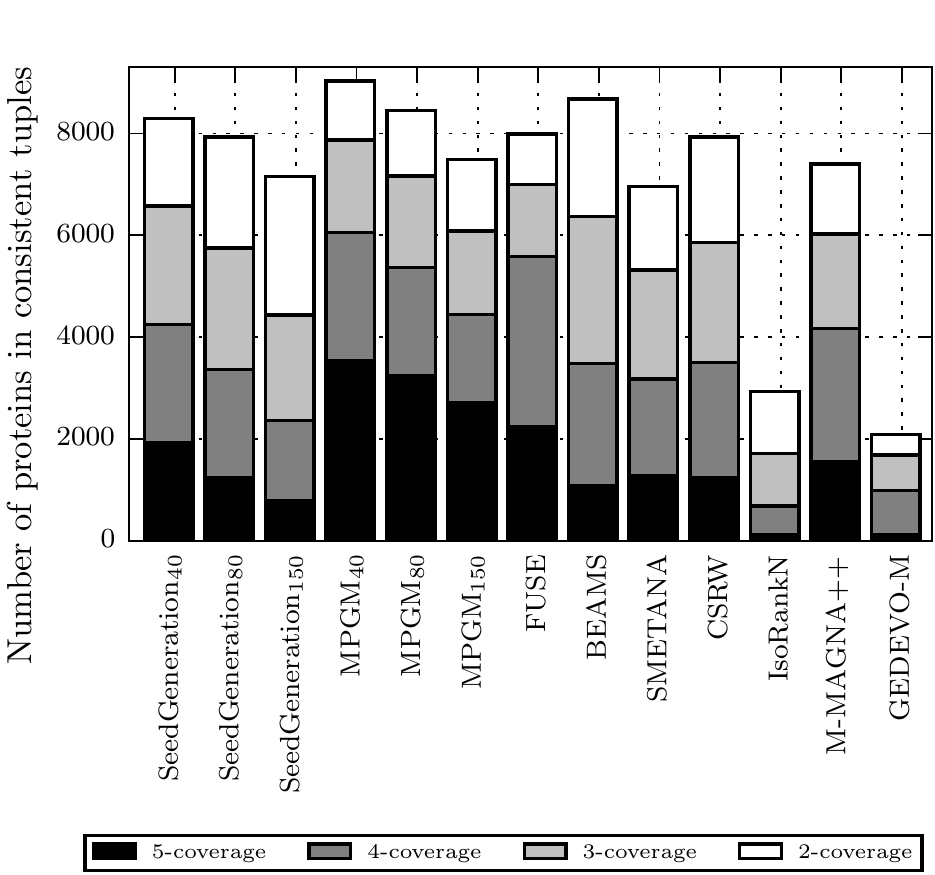}
	\caption{Number of proteins in consistent tuples with different 
		$d$-coverages. The results are for tuples with 
		nodes from five, four, three and two networks. For \AlgMPROPER, 
		we set $r=1$ and $\ell \in \{ 40,80,150\}$. We observe that
		\AlgMPROPER finds the most number of proteins in consistent tuples and  
		consistent tuple with nodes from all the five networks.} 
	\label{fig:nb_dcp}
\end{figure}

\begin{figure}	[htb!]
	\centering
	\centering
	\includegraphics[width=0.7\textwidth]{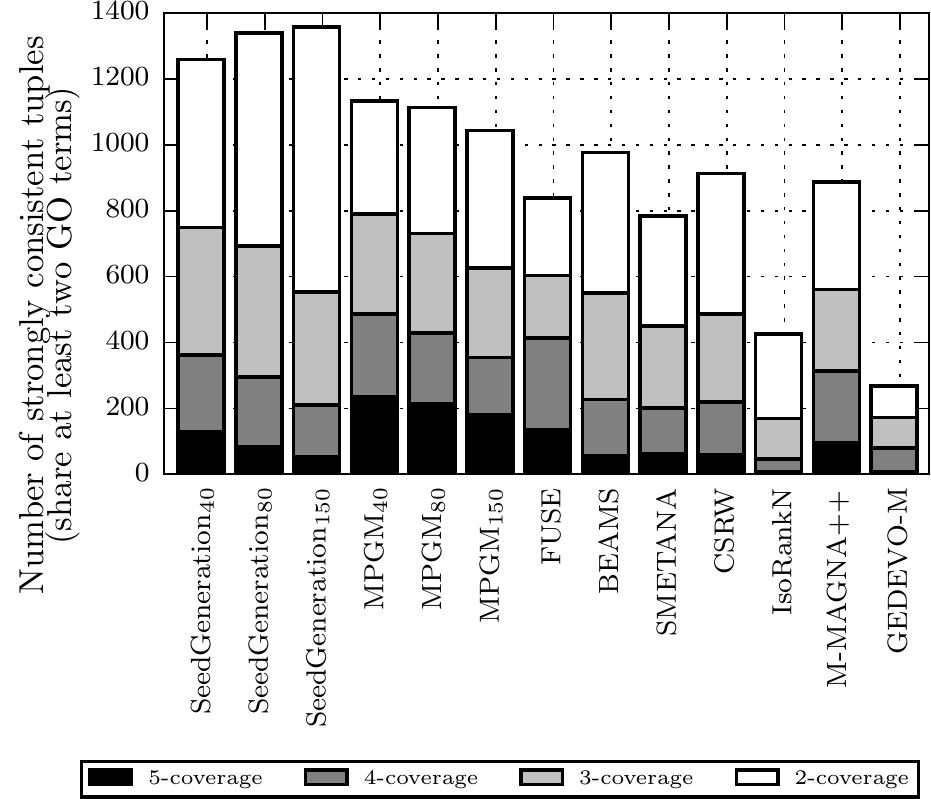}
	\caption{Number of strongly consistent tuples.}\label{fig:nb_c2}
\end{figure}

An algorithm with a good $d$-coverage does not necessarily generate 
high-quality tuples (in terms of functional similarity of proteins). For this 
reason, we look at the number of consistent tuples. For example, although 
IsoRankN generates the 
maximum number of tuples with proteins from two species (see 
Figure~\ref{fig:nb_m}), only a small 
fraction of these tuples are consistent (see Figure~\ref{fig:nb_c}). Also, in 
Figure~\ref{fig:nb_c}, 
we observe that \AlgMPROPER returns the largest number of consistent tuples with 
proteins from five different species. 
In addition, Figure~\ref{fig:nb_dcp} shows the number of proteins from consistent tuples with different $d$-coverages.
In order to further evaluate the functional similarities of aligned proteins, we also used a stronger notion for a consistent tuple: an annotated tuple is strongly consistent if all of the annotated proteins in that tuple share at least \textit{two} GO terms. In Figure~\ref{fig:nb_c2} we compare algorithms based on the number of strongly consistent tuples. We observe that \AlgSeed returns the most number of strongly consistent tuples. Also, \AlgMPROPER performs better than all the other state-of-the-art algorithms.

In \cref{table:specificity}, we compare specificity of algorithms.
Note that (i) chance of having better specificity for tuples of smaller sizes is higher, and (ii) different algorithms tend to output alignments with varying distribution of tuple sizes.
For this reason, we report specificities of tuples of size $5,4,3$ and $2$ separately.
We observe that  \AlgSeed provides the alignments with the best specificity. 
The main reason for this good performance is that it only used the sequence similarity information.
Also, the performance of \AlgMPROPER (in comparison to the other algorithms) is better for larger tuples.

\begin{table*}[t]
	\begin{center}
		\caption{Comparison results for  specificity of tuples of different sizes.
		 For \AlgMPROPER we set $r=1$.} \label{table:specificity}
	{\scriptsize
		\begin{tabular}{|l|ccc|ccc|c|c|c|c|c|c|c|}
			\hline
			& \multicolumn{3}{c|}{SeedGeneration ($\ell$)}  &  
			\multicolumn{3}{c|}{\AlgMPROPER ($\ell$)}   & F 
			&  B & S & C & I & M & G\\
			& 40 &  80 &  150  &  40 & 80 &  150 & & & & & & & \\
			\hline \hline
			$Spec_5$ &  \textbf{0.291} & 0.286 & 0.284 & 0.244 & 0.222 & 0.184 & 0.21 & 
			0.22 & 0.187 & 0.185 & 0.063 & 0.194 & 0.013 \\
			$Spec_4$ &  0.339 & 0.366 & \textbf{0.369} & 0.277 & 0.256 & 0.224 & 0.299 & 
			0.329 &0.291 & 0.306 & 0.092 & 0.297 & 0.075 \\
			$Spec_3$ &  0.462 & 0.486 & \textbf{0.500} & 0.384 & 0.382 & 0.349 & 0.35 & 
			0.437 &0.417 & 0.436 & 0.153 & 0.397 & 0.134 \\
			$Spec_2$ &  0.611 & 0.619 & \textbf{0.646} & 0.527 & 0.519 & 0.529 & 0.462 & 
			0.557 &  0.558 & 0.618 & 0.231 & 0.495 & 0.229  \\
			\hline
		\end{tabular}
	}
	\end{center}
\end{table*}

Tables~\ref{table:pgmcc5}, \ref{table:pgmcc4}, \ref{table:pgmcc3} and \ref{table:pgmcc2} (See Appendix~\ref{sec:tables}) provide detailed comparisons for tuples with different 
coverages.
More precisely, Table~\ref{table:pgmcc5} (Appendix~\ref{sec:tables}) compares algorithms over tuples with nodes from 
five networks. The second step of \AlgMPROPER (i.e., \AlgPercolate) 
uses PPI networks to generate 3076 tuples out of initial seed-tuples. We 
observe that \AlgMPROPER (for $\ell = 40$) finds an alignment 
with the maximum $d$-coverage, $\#CC_5$, $\#CP_5$ and $SS_p(\pi)$.
In addition, the first step of \AlgMPROPER (i.e., \AlgSeed) has the 
best performance on $Spec_5$,  $\overline{SS_p}(\pi)$ and MNE. This was 
expected, because \AlgPercolate uses only network structure, 
a less reliable source of information for functional similarity in 
comparison to sequence similarities, to align new nodes. From this table, it is 
clear that \AlgMPROPER outperforms the other algorithms with respect to all the 
measures.

\begin{figure}[htb!]
	\centering
	\includegraphics[width=0.8\textwidth]{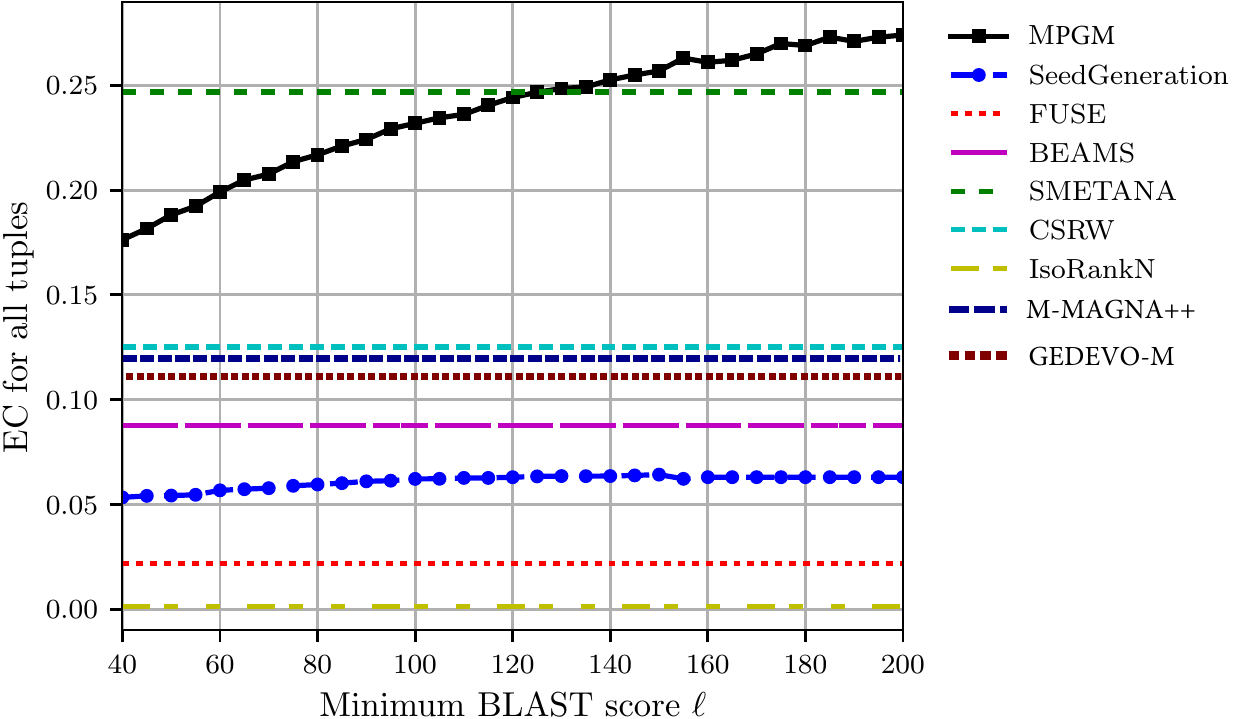}
	\caption{Comparison based on EC for all tuples. The 
		results for \AlgMPROPER are presented for $r=1$ and different values of $\ell$.} 
	\label{fig:ECt}
\end{figure}

\begin{figure}[htb!]
	\centering
	\includegraphics[width=0.8\textwidth,keepaspectratio]{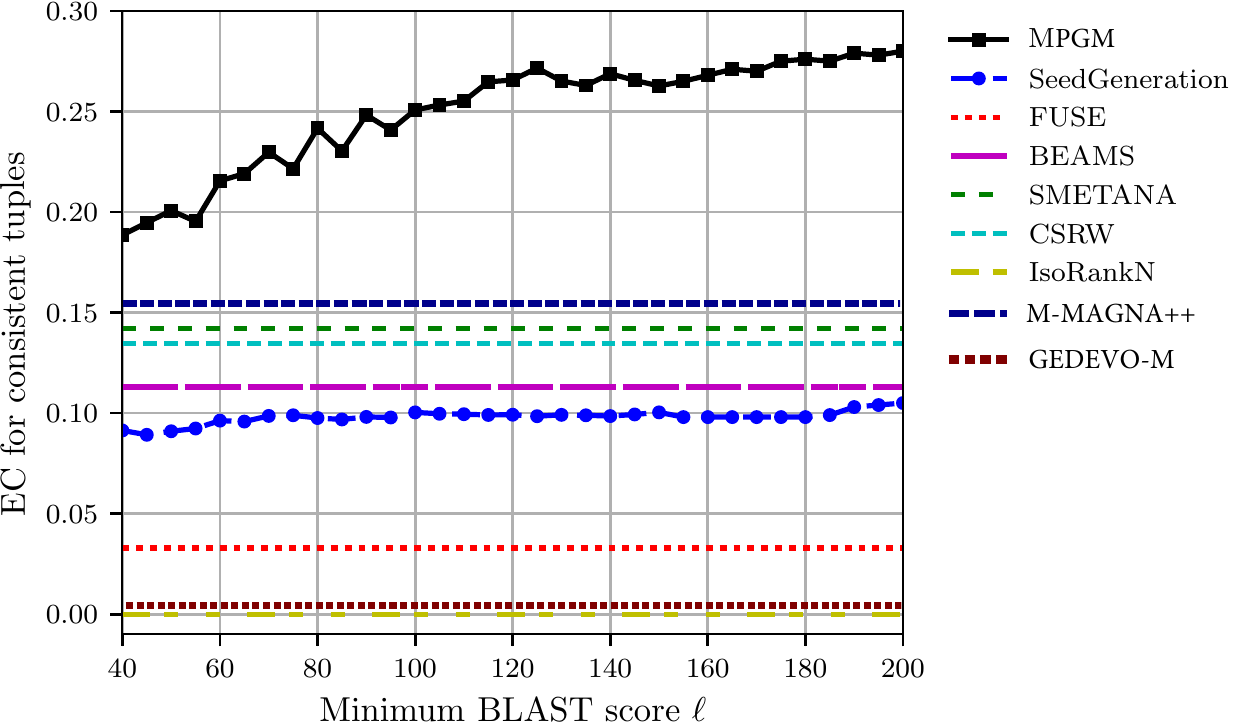}
	\caption{Comparison based on the EC measure for consistent tuples. The 
		results for \AlgMPROPER are presented for $r=1$ and different values of $\ell$. We 
		observe that \AlgMPROPER has the best performance among all 
		the algorithms.} 
	\label{fig:ECc}
\end{figure}

\begin{figure}[htb!]
	\centering
	\includegraphics[width=0.8\textwidth, keepaspectratio]{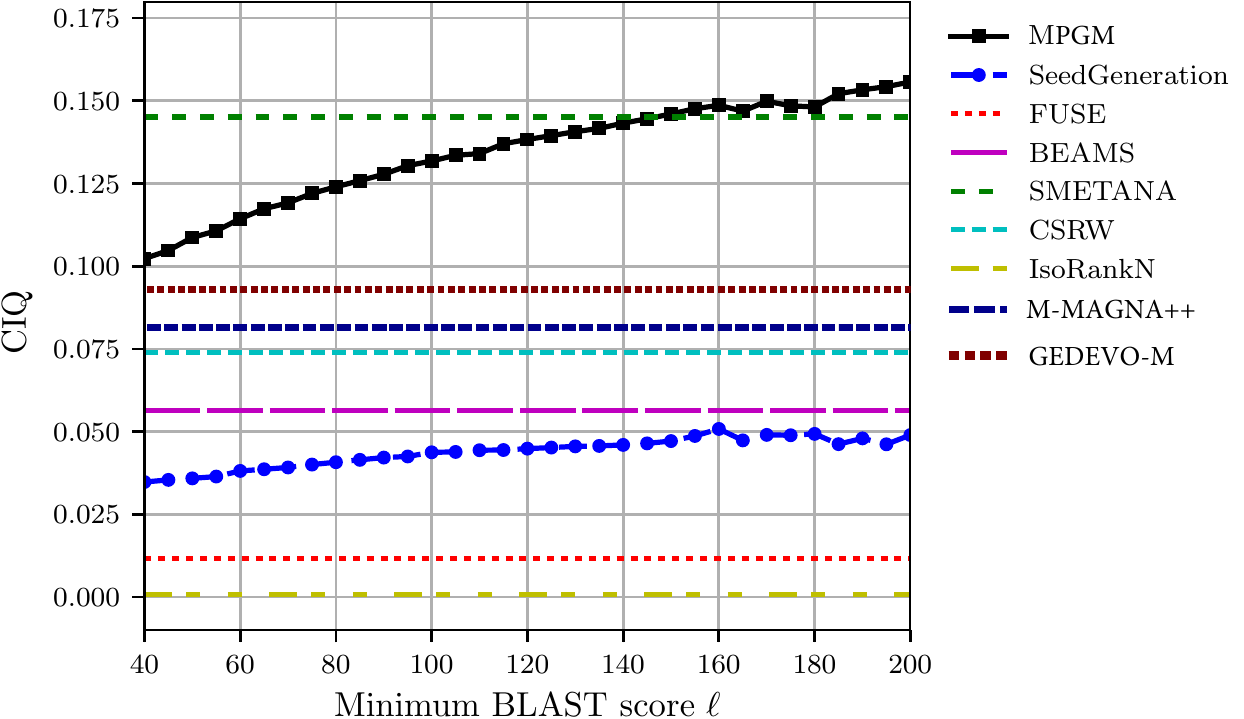}
	\caption{Comparison based on the CIQ measure. The results for \AlgMPROPER are 
		presented for $r=1$ and different values of $\ell$. We 
		observe that \AlgMPROPER and SMETANA find 
		alignments with the highest CIQ score.} 
	\label{fig:CIQ}
\end{figure}

Figure~\ref{fig:ECt} compares algorithms based on the EC measure. We 
observe that \AlgMPROPER (for values of $\ell$ larger than 150) finds
alignments with the highest EC score.
In Figure~\ref{fig:ECc}, to calculate EC, we consider only the edges between 
consistent tuples. We observe that \AlgMPROPER has the best performance among all 
the algorithms. This shows that \AlgMPROPER finds alignments where (i) many of 
the aligned tuples are consistent and (ii) there are many conserved 
interactions 
among these consistent tuples. CIQ is another measure, based on the structural 
similarity of aligned networks, for further evaluating
the performance of algorithms.
In Figure~\ref{fig:CIQ}, we observe that \AlgMPROPER and SMETANA find 
alignments with the best CIQ score.

To sum-up, we observe that \AlgMPROPER generally provides a nice trade-off between functional and structural similarity measures which is the Pareto frontier. That is, one cannot choose an algorithm that does better on both measures. Finally, our recommend setting for \AlgMPROPER is $r=1$ and $\ell = 80$.

\subsection{Computational Complexity}
The computational complexity of the \AlgSeed algorithm is 
$O\left(|\mathcal{S}_{\geq \ell}| \log{|\mathcal{S}_{\geq \ell}|}\right)$; 
it includes (i) sorting all the sequence similarities from the 
highest to the lowest, and (ii) processing them.
The computational complexity of the \AlgPercolate algorithm is 
$O\left(k^2\left(|E_1| + 
|E_2|\right) \min \left(D_1, 
D_2\right)\right)$, where $D_{1,2}$ are the maximum degrees in the two networks.

One of the key features of our algorithm is its computational simplicity. \AlgPercolate is easily parallelizable in a distributed manner. 
To have a scalable algorithm, for very large networks (graphs with millions of nodes), we use a MapReduce \cite{dean2008mapreduce} implementation of \AlgPercolate. 
The MapReduce programming model consists of two important steps \cite{dean2008mapreduce}: First, the Map step processes a subset of data (based on the task) and returns another set of data. Second, the Reduce  step, from the result of the Map step, returns a smaller set of data. 
More specifically, for \AlgPercolate, in the Map step, we spread out marks from the aligned proteins. In the Reduce step, we add all the couples with at least $r$ marks to the set of already aligned proteins $\pi$. The \AlgPercolate algorithm by  iteratively performing these two steps aligns the proteins from all the $k$ networks.

In \cref{table:time}, we compare the time complexity of different algorithms in order to align all the five species form \cref{table:ppi-1}. 
We observe that \AlgMPROPER and SMETANA are the fastest algorithms.
Also, while FUSE shows the closest performance to \AlgMPROPER in terms of biological and topological measures, it is much slower.  
We also evaluated the performance of our MapReduce implementation in scenarios where the original implementation cannot perform the alignment process. For example, by using a Hadoop cluster of 25 nodes, it takes almost 27 minutes to align 5 synthetic networks with 5 million nodes.

\begin{table}[h]
	\begin{center}
		\caption{The time complexity (in seconds) of algorithms for
			aligning the five species from \cref{table:ppi-1}.} \label{table:time}
		\begin{tabular}{|l|r|}
			\hline
		Algorithm & Time (s)  \\
			\hline \hline
			\AlgMPROPER &  \textbf{131}\\
			FUSE & 5911 \\
			BEAMS &  364\\
			SMETANA &  146 \\
			CSRW &  626 \\
				IsoRankN & 9094 \\
			 multiMAGNA++ & 423\\ 			 
			 	  GEDEVO-M & 4746 \\
			\hline
		\end{tabular}
	\end{center}
\end{table}

\section{Interpretation and Discussion}\label{sec:rationale}
One simple solution to the global MNA problem is to first 
compute individual alignments between all pairs of networks and then
derive the final multiple 
alignment by merging all these pairwise alignments. The main drawback of this 
approach is that the collection of these pairwise alignments might be 
inconsistent. For example, for nodes $u_{1,2,3} \in V_{1,2,3}$, if $u_1$ is 
matched to $u_2$ and $u_2$ to $u_3$, but $u_1$ is matched to another node from 
$G_3$, then it is not possible to generate a consistent one-to-one global MNA 
from these pairwise alignments.
In contrast to the idea of merging different pairwise alignments, our approach 
has three main advantages:
(i) It aligns all the $k$ networks at the same time. Therefore, it will always end up with a consistent one-to-one global MNA. (ii) It uses structural information from all networks simultaneously. (iii) The \AlgSeed algorithm gives more weight 
to the pairs of species that are evolutionarily closer to each other. For example, as H. sapiens and M. 
musculus are  very close, (a) many couples from these two species are matched first, and (b) there are more couples of 
proteins with high sequence similarities from these two species. Hence there are more tuples that contain proteins from both H. sapiens and M. musculus. 
In the rest of this section, we provide experimental evidence and theoretical results that explain the good performance of the \AlgMPROPER algorithm. 


\subsection{Why Does SeedGeneration Work?}  \label{sec:whyseedgen}
The first step of \AlgMPROPER (\AlgSeed) is a heuristic 
algorithm that generates seed-tuples. 
The \AlgSeed algorithm is designed based on the following observations.
First, it is well known that proteins with high BLAST bit-score similarities share GO terms with a high probability.\footnote{For a detailed discussion on this argument please refer to Appendix~\ref{go_sim}.}
Second, we look at the transitivity of BLAST bit-score similarities for real proteins.
Note that the BLAST similarity, in general, is not a transitive measure, i.e., for proteins $p_1, p_2$ and $p_3$ given that couples 
$[p_1, p_2]$ and $[p_2, p_3]$ are similar, we can not always conclude that the two proteins $p_1$ and $p_3$ are similar (see Example~\ref{example:trans}). 
\begin{example} \label{example:trans}
	Consider the three toy proteins $p_1, 
	p_2$ and $p_3$ with amino-acid sequences $[MMMMMM]$, $[MMMMMMVVVVVV]$ and 
	$[VVVVVV]$, respectively. In this example, $p_2$ is similar to both $p_1$ and 
	$p_3$, where $p_1$ is not similar to $p_3$. Indeed, we have $BlastBit(p_1 , 
	p_2) = 11.2, BlastBit(p_2 , p_3 ) = 10.0$ and $BlastBit(p_1 , p_3) = 0$. 
\end{example}
In real-world, proteins cover a small portion of the space of possible amino-acid 
sequences, and it might be safe to assume a (pseudo) transitivity property for 
them.
To empirically evaluate the transitivity of BLAST bit-scores, we define a new 
measure for an estimation of the BLAST bit-score similarity of two 
proteins $p_1$ and $p_3$, when we know that there is a protein $p_2$, such that 
BLAST bit-score similarities between $p_2$ 
and both $p_1,p_3$ are at least $\ell$. Formally, we define 
$\alpha_{\ell, \beta}$ as
\begin{align*}
	\alpha_{\ell, \beta} = \argmax_{\alpha}  [  \mathbb{P}   [BLAST&(i,k)    \geq \alpha \times \ell \ |  \ BLAST(i,j) \\ &\geq \ell, BLAST(j,k) \geq \ell ]   \geq \beta ] 
\end{align*}

\begin{figure}[ht!]
	\centering
	\includegraphics[width=0.8\textwidth]{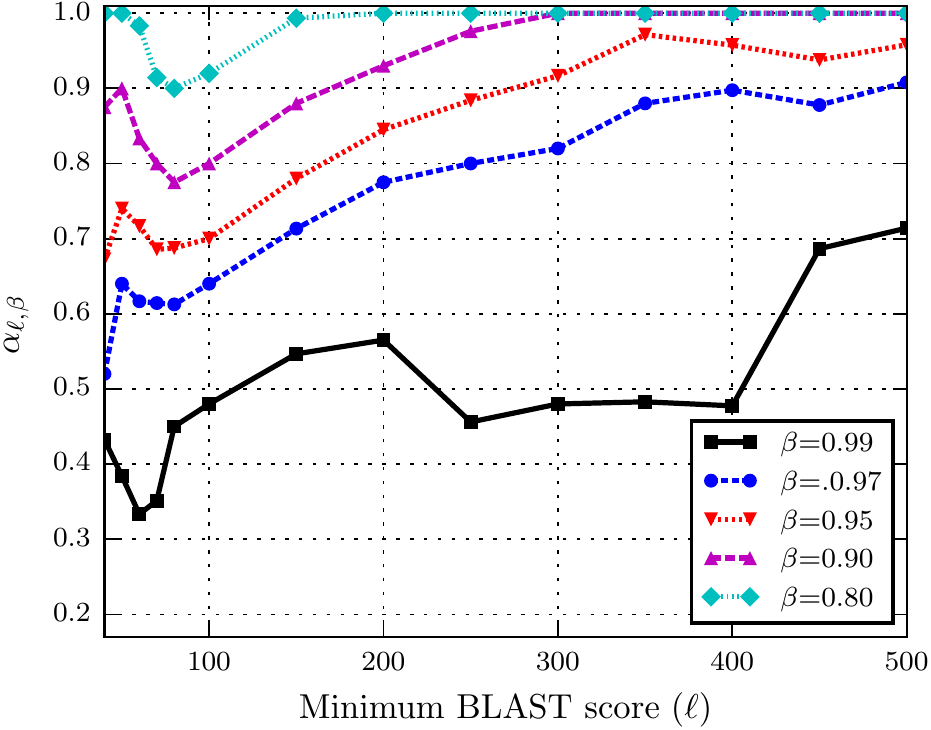}
	\caption{The transitivity of BLAST bit-score similarities for real proteins. 
		The $\alpha_{\ell, \beta}$ measure is calculated for different values of 
		$\ell$ and $\beta$.} 
	\label{fig:trans}
\end{figure}

An empirical value of $\alpha_{\ell, \beta}$\ close to one
is an indicator of a high level of 
transitivity (with a probability of $\beta$) between the sequence similarities 
of protein couples. 
In Figure~\ref{fig:trans}, we study the transitivity of BLAST bit-scores for 
different levels of confidence $\beta$. 
For example, in this figure, we observe that for two couples $[p_1,p_2]$ and 
$[p_2,p_3]$ with BLAST bit-score similarities of at least $100$, the similarity 
of the couple $[p_1,p_3]$ is at least 91 with a probability of $0.80$. In 
general, based on this experimental evidence, it seems reasonable to assume 
that there is a \textit{pseudo-transitive} relationship between the sequence similarities of real proteins.

The two main observations about (i) the relationship between sequence 
similarity and biological functions of protein couples, and (ii) 
the transitivity 
of BLAST bit-scores help us to design a heuristic algorithm for generating 
high-quality tuples (i.e., $\ell^*$-consistent tuples with value of $\ell^*$ very close to $\ell$) from sequence 
similarities. 


\subsection{Why Does MultiplePercolation Work?} 
\label{sec:whypercolationgm}

The general class of PGM algorithms has been shown to be very powerful for 
global pairwise-network alignment problems. For example, PROPER is a
state-of-the-art algorithm
that uses PGM-based methods to align two networks \cite{kazemi2016proper}. There 
are several works 
on 
the theoretical and practical aspects of PGM algorithms \cite{narayanan:2009, Yartseva:2013, korula14efficient, chiasserini2014anonymizing, kazemi2015when, kazemi2016network, cullina2016improved, cullina2016simultaneous, shiraniGE17, shiraniGE18, dai2018performance, mossel2018seeded}. In this paper, we introduced a global MNA algorithm, as a new  member of the PGM class.
In this section, by using a parsimonious $k$-graph sampling model (as a 
generalization of the model from \cite{kazemi2015when}), we prove that 
\AlgPercolate aligns all the nodes correctly if initially 
enough number of seed-tuples are provided. We first explain the model. Then we 
state the main theorem. Finally, we present experimental evaluations of 
\AlgPercolate over random graphs that are generated based on 
our $k$-graph sampling model.

\subsubsection{A Multi-graph Sampling Model}

Assume that all the $k$ networks $G_i(V_i,E_i)$ are evolved from an ancestor 
network $G(V,E)$ through node sampling (to model gene or protein deletion) and 
edge sampling (to model loss of protein-protein interactions) processes.

\begin{definition}[The $Multi(G,\boldsymbol{t},\boldsymbol{s},k)$ sampling 	model]
	Assume we have $\boldsymbol{t} = [t_1,t_2, \cdots, t_k]$ and $\boldsymbol{s} = 
	[s_1,s_2, \cdots, s_k]$, $0 < t_i, s_i \leq 1$. The network $G_i(V_i, E_i)$ is 
	sampled from $G(V,E)$ 
	in the following way: First the nodes $V_{i}$ are sampled from $V$ 
	independently with probability $t_i$; then the edges $E_i$ are 
	sampled from those edges of graph $G$, whose both endpoints are sampled in 
	$V_i$, 
	by independent edge sampling processes with probability $s_i$. We define 
	$t_{i,j} = \sqrt{t_it_j}$ and $s_{i,j} = \sqrt{s_is_j}$.
\end{definition}

\begin{definition} [A correctly matched tuple]
	A tuple $T$ is a correctly matched tuple, if and only if all the nodes in $T$ 
	are the same (say a node $u$), i.e., they are samples of a same node from the 
	ancestor network $G$.
\end{definition}

\begin{definition} [A completely correctly matched tuple]
	A correctly matched tuple $T$, which contains a different sample of a node $u$, 
	is complete if and only if for all the vertex sets $V_i, 1 \leq 
	i \leq k$, if $u \in V_i$ then $V_{i}(T) = u$
\end{definition}

Assume the $k$ networks $G_i(V_i,E_i)$ are sampled from a $G(n, p)$ random 
graph with $n$ nodes and average degrees of $np$. 
Now we state two main theorems that guarantee the performance of 
\AlgPercolate over the
$Multi(G(n.p),\boldsymbol{t},\boldsymbol{s},k)$ sampling model.
We first define two parameters 
$b_{t,s,r}$ and $a_{t,s,r}$:
\begin{align}  \label{threshold}
b_{t,s,r} = \left[\dfrac{(r-1)!}{nt^2(ps^2)^r}\right]^{\frac{1}{r-1}} \mbox{ and } 
a_{t,s,r} = (1 - \dfrac{1}{r})b_{t,s,r}. 
\end{align}

\begin{theorem} \label{theory:pgm} 
	For $r \geq 2$ and an arbitrarily small but fixed 
	$\tfrac{1}{6} > \epsilon > 0$, 
	assume that $ n^{-1} \ll p \leq n^{-\frac{5}{6} - \epsilon}$. For an initial 
	set of seed tuple $\mathcal{A}$, if 
	$|\mathcal{A}_{i,j}| > (1 + \epsilon)a_{t_{i,j},s_{i,j},r}$ for every $1 
	\leq i,j \leq k, i \neq j$, then with high 
	probability the \AlgPercolate algorithm percolates and for the 
	final alignment $\pi$, we 
	have $|\pi_{i,j}| = nt_{i,j}^2 \pm o(n)$, where almost all the tuples 
	are completely correctly matched tuples.
\end{theorem}

\begin{theorem} \label{theory:pgm_general}  
	For $r \geq 2$ and an arbitrarily small but fixed 
	$\tfrac{1}{6} > \epsilon > 0$, 
	assume that $ n^{-1} \ll p \leq n^{-\frac{5}{6} - \epsilon}$. For an initial 
	set of seed tuple $\mathcal{A}$, if for every $1 
	\leq i \leq k$ there at least $c$ set of $\mathcal{A}_{i,j}$, $1 \leq j \leq 
	k$ and $i \neq j$, such that
	$|\mathcal{A}_{i,j}| > (1 + \epsilon)a_{t_{i,j},s_{i,j},r}$,
	then with high 
	probability the \AlgPercolate algorithm percolates and for the 
	final alignment $\pi$, we 
	have:
	\begin{itemize}
		\item Almost all the tuples $T \in \pi$ are correctly matched tuples.
		\item For a correctly matched tuple $T$, which contains the node $u$, if there 
		are at least $k-c+1$ networks $G_i(V_i,E_1)$ such that $u \in V_i$, then $T$ is 
		a completely correctly matched tuple
	\end{itemize}
	
\end{theorem}
Note that Theorem~\ref{theory:pgm} is the special case of 
Theorem~\ref{theory:pgm_general} for $c = k-1$.
The proofs of Theorems~\ref{theory:pgm} and \ref{theory:pgm_general} follow 
from the generalization of the ideas that are used to prove
\cite[Theorem~1 (Robustness of {\textsc{NoisySeeds}}\xspace)]{kazemi2015growing}.

\subsubsection{Experimental Results: Synthetic Networks}
To evaluate the performance of our algorithm by using synthetic networks, we 
consider $k \in \{3,4,5\}$ randomly generated networks from the $Multi(G,\boldsymbol{t},\boldsymbol{s},k)$ model. 
In these experiments, we assume that a priori a set of seed-tuples $\mathcal{A}$
$(|\mathcal{A}| = a)$, with nodes from all the $k$ networks, are given and the 
\AlgPercolate  algorithm starts the alignment process from these tuples. 

\begin{figure}[!tpb]
	\centering
	\includegraphics[width=0.75\textwidth, keepaspectratio]{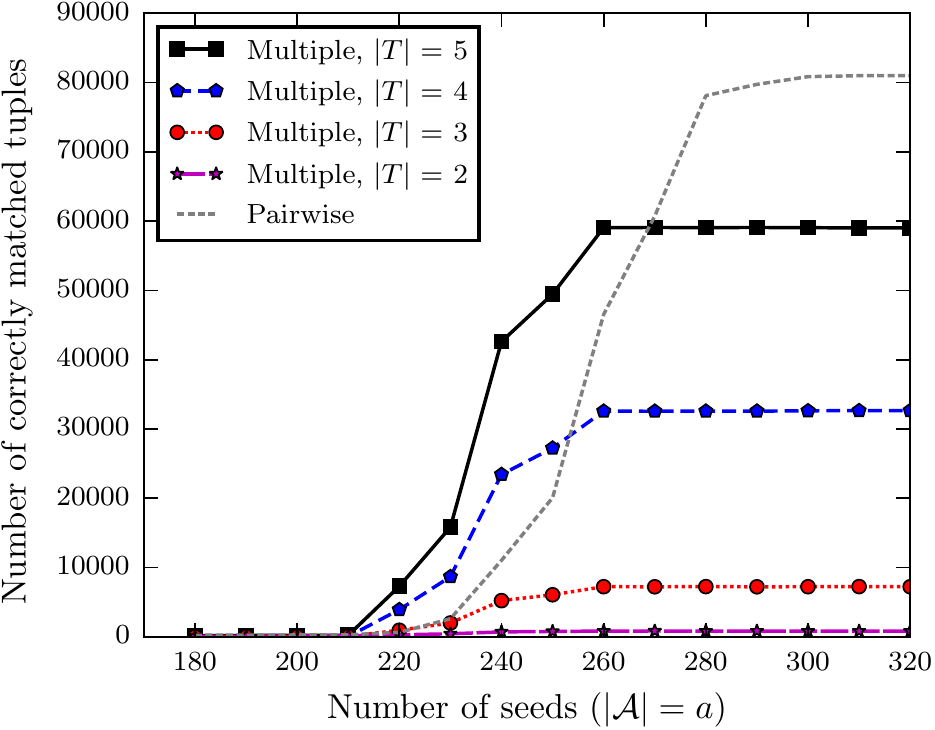}
	\caption{Multiple network alignment for graphs sampled from 
		$Multi(G,\boldsymbol{t},\boldsymbol{s},k)$ with parameters $k = 5, n = 
		10^5, p = 20/n, t = 0.9$ and $s = 0.9$. We set $r=2$ for 
		\AlgPercolate.} 
	\label{fig:pgmk_5_0.9_0.9}
\end{figure}

\begin{figure}[!tpb]
	\centering
	\includegraphics[width=0.75\textwidth, keepaspectratio]{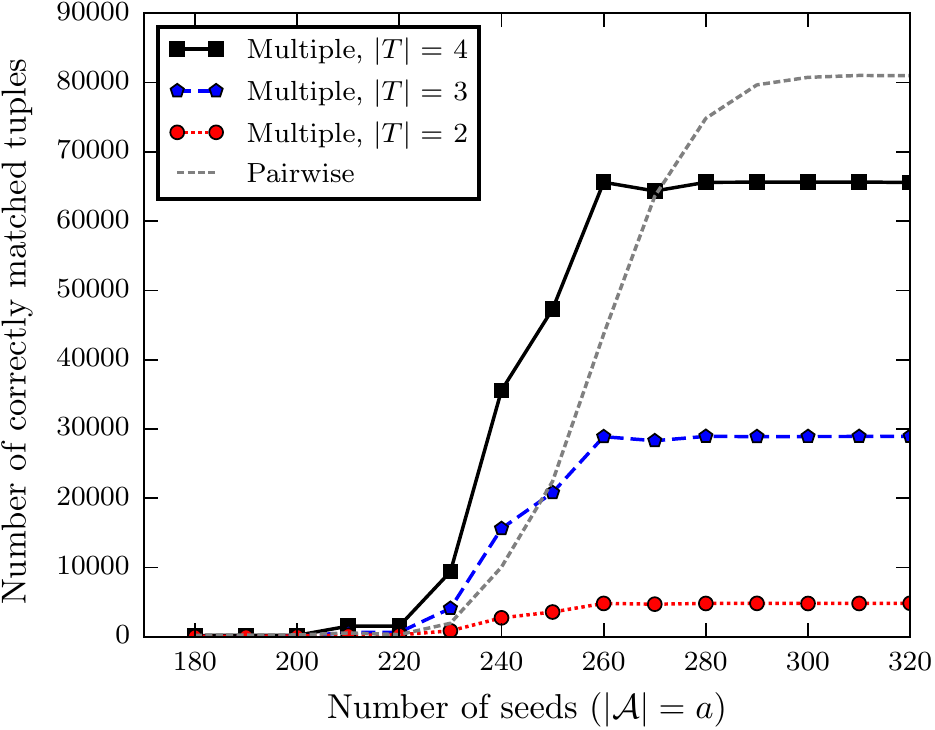}
	\caption{Multiple network alignment for graphs sampled from 
		$Multi(G,\boldsymbol{t},\boldsymbol{s},k)$ with parameters $k = 4, n = 
		10^5, p = 20/n, t = 0.9$ and $s = 0.9$. We set $r=2$ for 
		\AlgPercolate.} 
	\label{fig:pgmk_4_0.9_0.9}
\end{figure}

\begin{figure}[!tpb]
	\centering
	\includegraphics[width=0.75\textwidth, keepaspectratio]{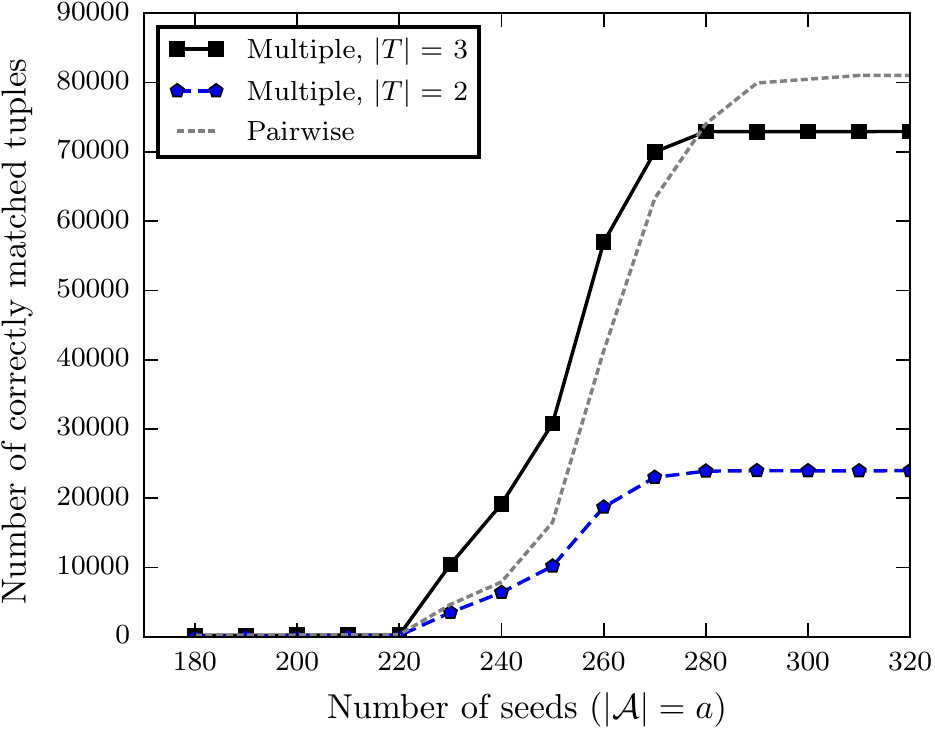}
	\caption{Multiple network alignment for graphs sampled from 
		$Multi(G,\boldsymbol{t},\boldsymbol{s},k)$ with parameters $k = 3, n = 
		10^5, p = 20/n, t = 0.9$ and $s = 0.9$. We set $r=2$ for 
		\AlgPercolate.} 
	\label{fig:pgmk_3_0.9_0.9}
\end{figure}

In the first set of experiments, we assume $G$ is an Erd\H{o}s-R\'{e}nyi graph with  $10^5$ nodes and an average degree of $20$. We assume $k$ networks $G_i$ are sampled from $G$ with node and edge sampling probabilities of $t=0.9$ and $s=0.9$, respectively.
Figures~\ref{fig:pgmk_5_0.9_0.9}, \ref{fig:pgmk_4_0.9_0.9} and \ref{fig:pgmk_3_0.9_0.9} show the simulation 
results for these experiments. We use $r=2$ for the 
\AlgPercolate 
algorithm.  For each $k \in \{3,4,5\}$, the total number of correctly aligned 
tuples is provided.
We observe that when there is enough number of tuples in the seed set, 
\AlgPercolate aligns correctly most of the nodes. We also see 
the sharp phase-transitions predicted in Theorems~\ref{theory:pgm} and 
\ref{theory:pgm_general}. According to Equation (2), we need 
$a_{t,s,r} = 236$ correct seed-tuples to find the complete alignments for 
the model parameters of $n = 10^5, p = 20/n, t = 0.9$ and $s = 0.9$.
We observe that the 
phase transitions take place very close to $a_{t,s,r} = 236$. For 
example, if $k =5$, in expectation there are $n t^{5} = 59049$ 
nodes that are present in all the five networks. From 
Figures~\ref{fig:pgmk_5_0.9_0.9} (the black curve), it is clear that 
\AlgPercolate aligns correctly almost all these nodes. Also, in 
expectation, there are $\binom{5}{3}nt^{3}(1-t)^2 = 7290$ nodes that are 
present in exactly three networks. Again, from Figures~\ref{fig:pgmk_5_0.9_0.9} 
(the red curve), we observe that \AlgPercolate correctly aligns 
them.

\begin{figure}[!tpb]
	\centering
	\includegraphics[width=0.75\textwidth, keepaspectratio]{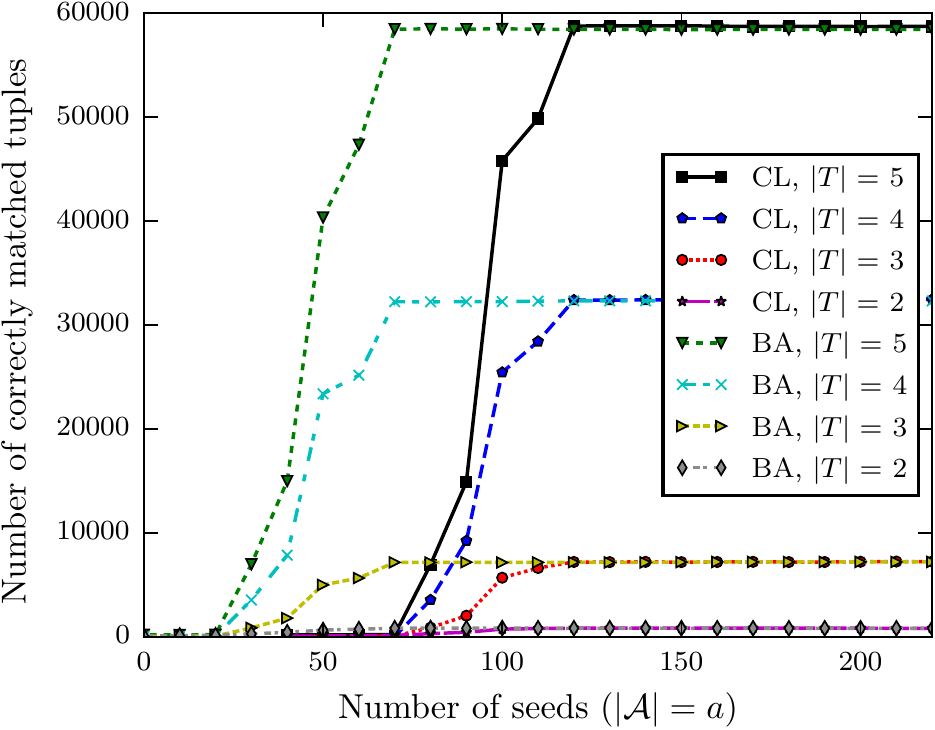}
	\caption{Multiple network alignment for graphs sampled from  Chung-Lu (CL) and Barab\'{a}si Albert (BA) models with $n= 10^5$ nodes.
		The average degree of both graphs are 10. The node ($t$) and edge  ($s$) sampling probabilities are both 0.9. We set $r=2$ for \AlgPercolate.} 
	\label{fig:pgmk-scalefree}
\end{figure}

While we are only able to guarantee the performance of the \AlgPercolate algorithm for Erd\H{o}s-R\'{e}nyi graphs, we study the performance of our algorithms on two other network models with heavy-tailed degree distributions. For this reason, first, we apply the \AlgPercolate algorithm to a variant of
power-law random graphs called the Chung-Lu model (CL) \cite{chung2002connected}. In this model, the degree distribution of nodes follows a power law.
Secondly, we apply  \AlgPercolate to the Barab\'{a}si Albert model (BA) \cite{barabasi1999emergence}. This model generates random scale-free networks in a preferential attachment setting.
In Figure~\ref{fig:pgmk-scalefree}, we observe that \AlgPercolate successfully aligns all the nodes from the $k=5$ networks correctly. We also observe that while for both models we need fewer number of seed than for Erd\H{o}s-R\'{e}nyi graphs, the required number of initial seeds for BA models is even less.

\section{Conclusion} \label{conclusion}
In this paper, we introduced a new one-to-one global multiple-network 
alignment algorithm, called 
\AlgMPROPER. Our algorithm has two main steps. In the first step 
(\AlgSeed), it uses protein sequence-similarities to 
generate an initial seed-set of tuples. In the second step, \AlgMPROPER applies a percolation-based 
graph-matching algorithm (called \AlgPercolate) to align the remaining unmatched proteins, by using only the 
structure of networks and the seed tuples from the first step. We have compared 
\AlgMPROPER with several state-of-the-art methods. We observe that \AlgMPROPER 
outperforms the other algorithms with respect to several measures. More 
specifically, \AlgMPROPER finds many consistent tuples with high $d$-coverage 
(mainly for $d=k$). Also, it outputs alignments with a high structural similarity between networks, i.e., many interactions are conserved among aligned tuples.
We have studied the transitivity of sequence similarities for real 
proteins and have found that it is reasonable to assume a \textit{pseudo-transitive}
relationship among them. We argue, based on this pseudo-transitivity property, 
that the\AlgSeed heuristic is able to find seed tuples with 
high functional similarities. In addition, we present a random-sampling model 
to generate $k$ correlated networks. By using this model, we prove conditions under which  \AlgPercolate aligns (almost) all the nodes correctly if initially enough seed tuples are provided. 
\paragraph{Acknowledgements.} The work of Ehsan Kazemi was supported by Swiss National Science Foundation (Early Postdoc.Mobility) under grant
number 168574.

\bibliographystyle{plainnat}
\bibliography{MNA}

\newpage 

\appendix

\section{Table of Notations} \label{sec-appendixA}
%

\begin{table}[ht!] 
\begin{center}
\caption{}\label{tableofnotations}\label{table:notation}
\begin{tabular}{|l|p{0.8\linewidth}|}
   \hline
  $G_i(V_i , E_i)$ & A network with vertex set $V_i$ and edge set $E_i$.\\
  $(u,v)$ & An edge between nodes $u$ and $v$.\\
  $N_i(u)$ & The set of neighbors of node $u$ in $G_i$. \\
  $BlastBit(u, v)$ & BLAST bit-score similarity of two proteins $u$ and $v$ \\
  $[u, v]$ & A couple of proteins $u$ and $v$. \\
  $T$ & A tuple. \\
  $\mathcal{A}$ & Initial seed-tuples. \\
    $\pi$ & The final alignment. \\
  $|T|$ & Number of nodes in tuple $T$. \\
  $V(T)$ & The set of networks such that have a node in the tuple $T$. \\
  $V(u)$ & The network $V_i$ such that $u \in V_i$. \\
  $\mathcal{S}_{\geq \ell}$ & The set of all couples with BLAST bit-score 
similarities at least $\ell$. \\
$\mathcal{A}(u)$ & Returns the tuple $T \in \mathcal{A}$ such that $u \in T$. If there 
is no such tuple, we define $\mathcal{A}(u) = \emptyset$.  \\
 $E_{T_i, T_j}$ & The set of all the 
interactions between nodes from the two tuples $T_i$ and $T_j$, i.e., $E_{T_i, 
T_j} = \{ e= (u, v) | u \in T_i, v \in T_j \}$.\\
 $V(E_{T_i, T_j})$ & The set of networks such that have an edge in $E_{T_i, 
T_j}$.\\
$C(\pi)$ & The set of consistent tuples in an alignment $\pi$. \\
  \hline
\end{tabular}
\end{center}
\end{table}
 
\section{Detailed Comparisons} \label{sec:tables}

Tables~\ref{table:pgmcc5}, \ref{table:pgmcc4}, \ref{table:pgmcc3} and \ref{table:pgmcc2} provide detailed comparisons for tuples with different coverages.
Table~\ref{table:pgmcc5} compares algorithms over tuples with nodes from five networks. The second step of \AlgMPROPER (i.e., \AlgPercolate) 
uses PPI networks to generate 3076 tuples out of initial seed-tuples. We observe that \AlgMPROPER (for $\ell = 40$) finds an alignment with the maximum $d$-coverage, $\#CC_5$, $\#CP_5$ and $SS_p(\pi)$. In addition, the first step of \AlgMPROPER (i.e., \AlgSeed) has the best performance on $Spec_5$,  $\overline{SS_p}(\pi)$ and MNE. This was expected, because \AlgPercolate uses only network structure, a less reliable source of information for functional similarity in comparison to sequence similarities, to align new nodes. From this table, it is clear that \AlgMPROPER outperforms the other algorithms with respect to all the measures.

\begin{table*}[h!]
\begin{center}
\caption{Comparison results for tuples of size 
five. For \AlgMPROPER we set $r=1$.} \label{table:pgmcc5}
{\scriptsize
\begin{tabular}{|l|ccc|ccc|c|c|c|c|c|c|c|}
	\hline
	& \multicolumn{3}{c|}{SeedGeneration ($\ell$)}  &  
	\multicolumn{3}{c|}{\AlgMPROPER ($\ell$)}   & F 
	&  B & S & C & I & M & G\\
	& 40 &  80 &  150  &  40 & 80 &  150 & & & & & & & \\
\hline \hline
$d$-coverage & 1366 & 880 & 568 & \textbf{3076} & 3062 & 3068 & 2233 & 867 & 
1132 & 1104 & 379 & 1896 & 1942 \\
$\#CC_5$ &  386 & 248 & 159 & \textbf{707} & 647 & 541 & 449 & 187 & 209 & 200 
& 23 & 312 & 25\\
$\#CP_5$ & 1930 & 1240 & 795 & \textbf{3535} & 3235 & 2705 & 2245 & 1082 & 1279 
& 1234 & 126 & 1560 & 125 \\
$Spec_5$ &  \textbf{0.291} & 0.286 & 0.284 & 0.244 & 0.222 & 0.184 & 0.21 & 
0.22 & 0.187 & 0.185 & 0.063 & 0.194 & 0.013 \\
$SS_p(\pi)$ &  5294 & 3519 & 2251 & \textbf{10659} & 10002 & 9285 & 7078 & 2788 
& 3315 & 3097 & 554 & 5279& 2614\\
$\overline{SS_p}(\pi)$ & 3.928 & \textbf{4.018} & 3.993 & 3.55 & 3.326 & 3.074 
& 3.22 & 3.239 & 2.944 & 2.818 & 1.482 &  3.071 & 1.37 \\
MNE & \textbf{2.927} & 2.943 & 3.049 & 3.008 & 3.071 & 3.144 & 3.014 & 3.162 & 
3.312 & 3.071 & 3.469 & 3.185 & 3.889\\
\hline
\end{tabular}
}
\end{center}
\end{table*}

\begin{table*}[h!]
\begin{center}
\caption{Comparison results for tuples of size four. For \AlgMPROPER 
we set $r=1$.}\label{table:pgmcc4}
{\scriptsize
\begin{tabular}{|l|ccc|ccc|c|c|c|c|c|c|c|}
	\hline
	& \multicolumn{3}{c|}{SeedGeneration ($\ell$)}  &  
	\multicolumn{3}{c|}{\AlgMPROPER ($\ell$)}   & F 
	&  B & S & C & I & M & G\\
	& 40 &  80 &  150  &  40 & 80 &  150 & & & & & & & \\
\hline \hline
$d$-coverage & 1933 & 1591 & 1133 & 2719 & 2520 & 2321 &{3527} & 1663 & 
1547 & 1670 & 1475 & 3305 &  \textbf{4168} \\
$\#CC_4$ &  580 & 532 & 392 & 631 & 534 & 435 & \textbf{834} & 510 & 414 & 474 
& 118 & 652 & 215 \\
$\#CP_4$ & 2320 & 2128 & 1568 & 2524 & 2136 & 1740 & \textbf{3336} & 2398 & 
1903 & 2272 & 560 & 2608 & 860 \\
$Spec_4$ &  0.339 & 0.366 & \textbf{0.369} & 0.277 & 0.256 & 0.224 & 0.299 & 
0.329 &0.291 & 0.306 & 0.092 & 0.297 & 0.075 \\
$SS_p(\pi)$ &  8309 & 7335 & 5465 & 10449 & 9087 & 7902 & 12829 & 7043 
& 5863 & 6522 & 2953 & \textbf{12840} & 7988 \\
$\overline{SS_p}(\pi)$ & 4.591 & 4.814 & \textbf{4.982} & 4.213 & 3.984 & 3.726 
& 4.095 & 4.34 & 3.922 & 4.044 & 2.156 & 3.951 & 1.947 \\
MNE & \textbf{2.565} & 2.648 & 2.717 & 2.571 & 2.621 & 2.668 & 2.597 & 2.733 & 
2.664 & 2.73 & 3.168 & 2.720 & 3.525\\
\hline
\end{tabular}
}
\end{center}
\end{table*}

\begin{table*}[h!]
\begin{center}
\caption{Comparison results for tuples of size three. For \AlgMPROPER 
we set $r=1$.}\label{table:pgmcc3}
{\scriptsize
\begin{tabular}{|l|ccc|ccc|c|c|c|c|c|c|c|}
	\hline
	& \multicolumn{3}{c|}{SeedGeneration ($\ell$)}  &  
	\multicolumn{3}{c|}{\AlgMPROPER ($\ell$)}   & F 
	&  B & S & C & I & M & G\\
	& 40 &  80 &  150  &  40 & 80 &  150 & & & & & & & \\
\hline \hline
$d$-coverage & 2342 & 2227 & 1842 & 2502 & 2522 & 2523 & 2180 & 2320 & 1951 & 
1981 & \textbf{2869}  & 2736 & 2157 \\
$\#CC_3$ &  775 & 794 & 692 & 603 & 598 & 545 & 472 & \textbf{801} & 617 & 662 
& 308  & 620 & 232\\
$\#CP_3$ & 2325 & 2382 & 2076 & 1809 & 1794 & 1635 & 1416 & \textbf{2886} & 
2132 & 
2352 & 1027 & 1860 & 696 \\
$Spec_3$ &  0.462 & 0.486 & \textbf{0.500} & 0.384 & 0.382 & 0.349 & 0.35 & 
0.437 &0.417 & 0.436 & 0.153 & 0.397 & 0.134 \\
$SS_p(\pi)$ &  11509 &11672 & 9988 & 10040 & 10070 & 9430 & 8197 & 
11509 & 9064 & 9526 & 7463 & \textbf{14530} & 6092\\
$\overline{SS_p}(\pi)$ & 6.007 & 6.319 & \textbf{6.394} & 5.263 & 5.348 & 4.995 
& 4.956 & 5.706 & 5.441 & 5.587 & 3.224 & 5.389 & 2.884 \\
MNE & 2.239 & \textbf{2.290} & 2.372 & 2.312 & 2.336 & 2.35 & 2.348 & 2.276 & 
2.31 & 2.264 & 2.83 & 2.431 & 3.022\\
\hline
\end{tabular}
}
\end{center}
\end{table*}

\begin{table*}[h!]
\begin{center}
\caption{Comparison results for tuples of size two. For \AlgMPROPER 
we set $r=1$. }\label{table:pgmcc2}
{\scriptsize
\begin{tabular}{|l|ccc|ccc|c|c|c|c|c|c|c|}
\hline
& \multicolumn{3}{c|}{SeedGeneration ($\ell$)}  &  
\multicolumn{3}{c|}{\AlgMPROPER ($\ell$)}   & F 
&  B & S & C & I & M & G\\
& 40 &  80 &  150  &  40 & 80 &  150 & & & & & & & \\
\hline \hline
$d$-coverage & 3510 & 4013 & 4411 & 3402 & 3675 & 3825 & 3118 & 3265 & 2820 & 
2988 & \textbf{5620} & 3958 &  2110\\
$\#CC_2$ &  859 & 1088 & \textbf{1357} & 579 & 645 & 703 & 495 & 900 & 702 & 
905 & 541 & 685 & 202 \\
$\#CP_2$ & 1718 & 2176 & \textbf{2714} & 1158 & 1290 & 1406 & 990 & 2309 & 1644 
&  2073 & 1224 & 1370 & 404 \\
$Spec_2$ &  0.611 & 0.619 & \textbf{0.646} & 0.527 & 0.519 & 0.529 & 0.462 & 
0.557 &  0.558 & 0.618 & 0.231 & 0.495 & 0.229  \\
$SS_p(\pi)$ &  15049 & 18777 & \textbf{22849} & 10664 & 12032 & 12918 & 9035 & 
14749 & 12003 & 14898 & 12853 & 22048 & 7897 \\
$\overline{SS_p}(\pi)$ & 8.157 & 8.286 & \textbf{8.551} & 7.025 & 7.153 & 7.161 
& 6.118 & 7.375 & 7.378 & 8.124 & 4.21 & 5.729  & 3.826\\
MNE & 1.946 & \textbf{1.944} & 1.968 & 2.01 & 2.03 & 2.002 & 2.163 & 1.987 & 
1.951 & 1.987 & 2.464 & 2.152 & 2.797 \\
\hline
\end{tabular}
}
\end{center}
\end{table*}

 \label{detail-table}
\section{BLAST Bit-score Similarities and GO terms} \label{go_sim}

\begin{figure}[!tpb]
	\centering
	\includegraphics[width=10cm, keepaspectratio]{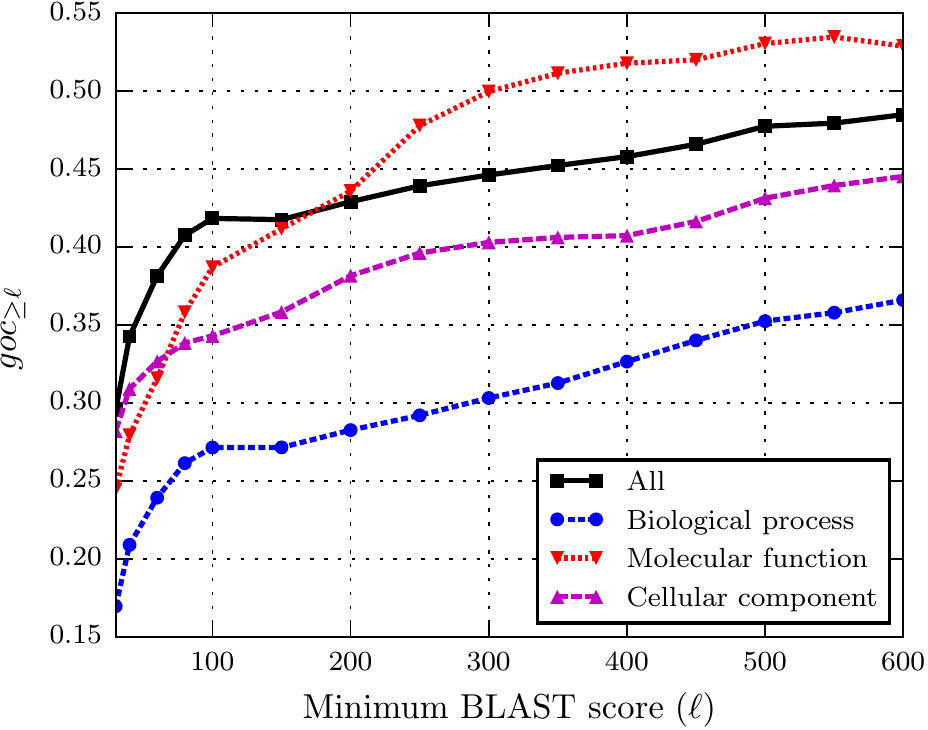}
	\caption{The $goc_{\geq \ell}$ measure for couples of proteins with BLAST 
		bit-score similarities of at least $\ell$.} 
	\label{fig:blast_go}
\end{figure}

\begin{figure}[!tpb]
	\centering
	\includegraphics[width=10cm, keepaspectratio]{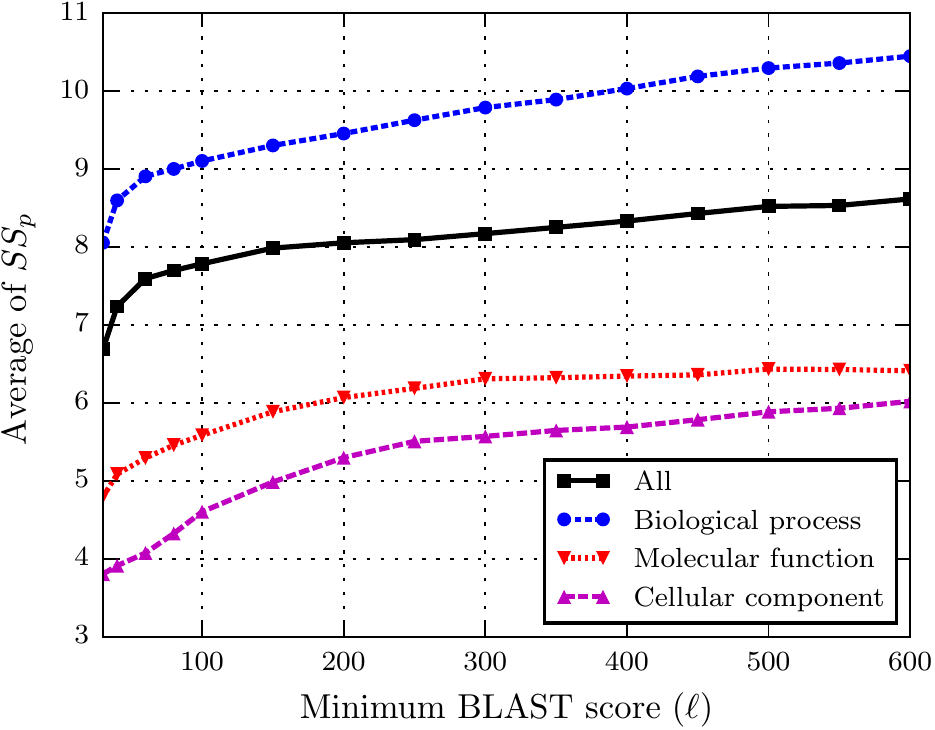}
	\caption{Average of $SS_p$ for couples of proteins with BLAST bit-score 
		similarities of at least $\ell$. We observe that, for couples 
		of proteins with higher BLAST bit-score similarities, the average of $SS_p$ 
		measure increases.} 
	\label{fig:blast_Resnik}
\end{figure}

To provide experimental evidence for our hypothesis, we look at the biological 
similarity of protein couples versus their BLAST bit-score similarities.
For this reason, we define a gene-ontology consistency (GOC) measure (based on 
the measure introduced in \cite{mistry2008gene}) to evaluate the 
relationship 
between BLAST bit-scores and the experimentally verified GO terms.
This measure represents the percentage of pairs of proteins with BLAST bit-score
similarity of at least $\ell$, such that they share at least one GO term. 
Formally, we define
\begin{align}
& goc_{\geq \ell} =  \nonumber \\ & \dfrac{|\{ [p_i, p_j] | BLAST(p_i, p_j) \geq \ell \text{ and 
	} 
	go(p_i) \cap go(p_j) \neq \emptyset \}|}{|\{ [p_i, p_j] | BLAST(p_i, p_j) \geq 
	\ell, go(p_i) \neq \emptyset \text{ and } 
	go(p_j) \neq \emptyset \}|}.
\end{align}
In this section, we consider only experimentally verified GO 
terms. 
Figure~\ref{fig:blast_go} shows the $goc_{\geq \ell}$ measure for couples of 
proteins among five eukaryotic species, namely C. elegans (ce), D. melanogaster 
(dm), H. sapiens (hs), M. musculus (mm) and S. cerevisiae (sc).
In this figure, the results are provided for cases, where we consider (i) all 
the experimental GO terms, (ii) cellular component (CC) annotations, 
(iii) molecular function (MF) annotations, and (iv) biological process (BP) 
annotations.
For further experiments, we look at the average of semantic similarity $SS_p$ (\cref{eq:ss_p}) between couples of proteins with BLAST bit-score similarity of 
at least $\ell$. Figure~\ref{fig:blast_Resnik} shows the $SS_p$ for couples  of 
proteins with 
BLAST bit-score similarities of at least $\ell$. We observe that, for couples 
of proteins with higher BLAST bit-score similarities, the average of $SS_p$ 
measure increases.
\section{GO Annotation: Statistics} \label{go_stat}
In this appendix, we look at a few statistics regarding GO annotations.
GO annotations comprises three orthogonal taxonomies for a gene product: 
molecular-function, biological-process 
and  cellular-component. This information is captured in 
three different directed acyclic graphs (DAGs). The roots (the most general 
annotations for each category) of these DAGs are:
\begin{itemize}
 \item GO:0003674 for molecular function annotations
 \item GO:0008150 for biological process annotations
 \item GO:0005575 for cellular component annotations
\end{itemize}

For information content of each GO term, we use the 
SWISS-PROT-Human proteins, and counted the number of times each concept occurs. 
Information content is calculated based on the following information:
\begin{itemize}
 \item Number of GO terms in the dataset is 26831.
 \item Number of annotated proteins in the dataset is 38264085.
  \item Number of experimental GO terms in the dataset is 24017.
 \item Number of experimentally annotated proteins in the dataset is 102499.
\end{itemize}

Table~\ref{table:ppi-information} provides information related to different 
categories of GO annotations for the five networks we used in our 
experiments.

\begin{table}[htb!]
\begin{center}
\caption{Statistics for experimental GO 
annotations.}\label{table:ppi-information}
\begin{tabular}{|l|r|r|r|}
\hline
GO type & $\#$GO & $\#$proteins & Avg. $\#$GO \\
\hline \hline
All &20738 & 28896 & 49.47 \\
Biological process & 14876 & 20723 & 48.21 \\
Molecular function & 3938 & 21670 & 7.84 \\
Cellular component & 1924 & 21099 & 12.35 \\ \hline
\end{tabular}
\end{center}
\end{table}

Next we report the number of experimentally annotated proteins (at the cut-off 
level 5 of DAGs) in each network:
\begin{itemize}
\item C. elegans: 1544 out of 4950 proteins (31.2 \%).
\item D. melanogaster: 4653 out of 8532 proteins (54.5 \%).
\item H. sapiens: 10929 out of 19141 proteins (57.1 \%).
\item M. musculus: 7150 out of 10765 proteins (66.4 \%).
\item S. cerevisiae: 4819 out of 6283 proteins (76.7 \%).
\end{itemize}
The probabilities of sharing at least one GO term (at the cut-off level 5) 
for tuples of size two to five, when all the proteins of a tuple are
annotated, are as follows:
\begin{itemize}
 \item tuples of size 2: 0.215
 \item tuples of size 3: 0.042 
 \item tuples of size 4: 0.009
 \item tuples of size 5: 0.002
\end{itemize}
Also, the probabilities of sharing at least one GO term (at the cut-off level 
5) for tuples of size two to five, when at least two proteins from each 
tuple are annotated, are as follows:
\begin{itemize}
  \item tuples of size 2: 0.215
  \item tuples of size 3: 0.167
  \item tuples of size 4: 0.120
  \item tuples of size 5: 0.081
\end{itemize}

\begin{figure}[htb!]
\begin{center} 
\includegraphics[width=10cm, keepaspectratio]{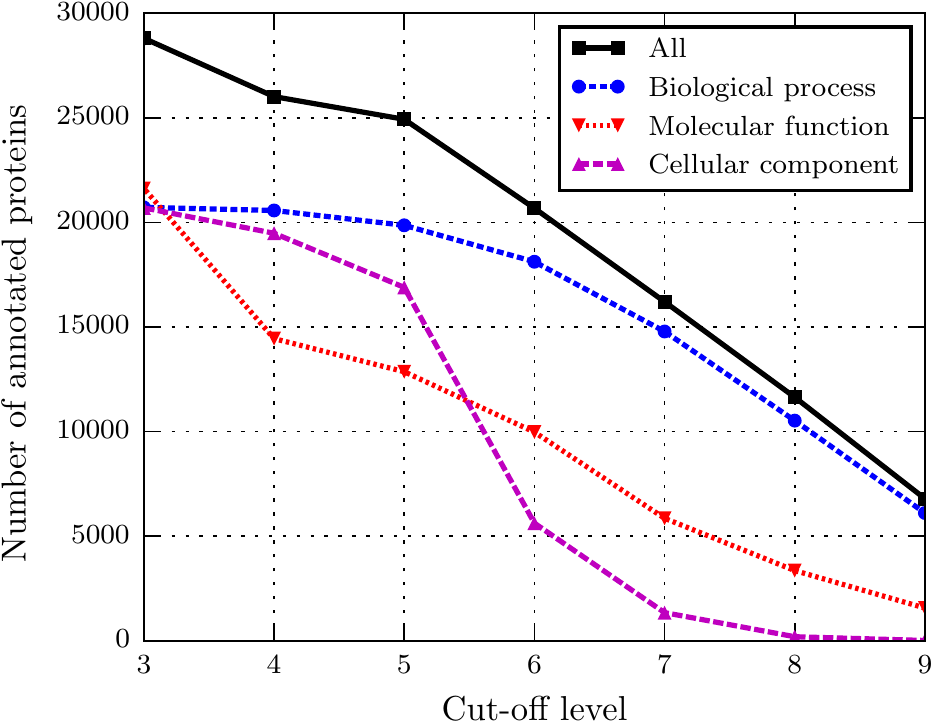}
  \caption{Number of annotated proteins for different cut-off levels} 
\label{fig:nb_prot_go_level}
\end{center}
\end{figure}

\begin{figure}[htb!]
\begin{center} 
\includegraphics[width=10cm,keepaspectratio]{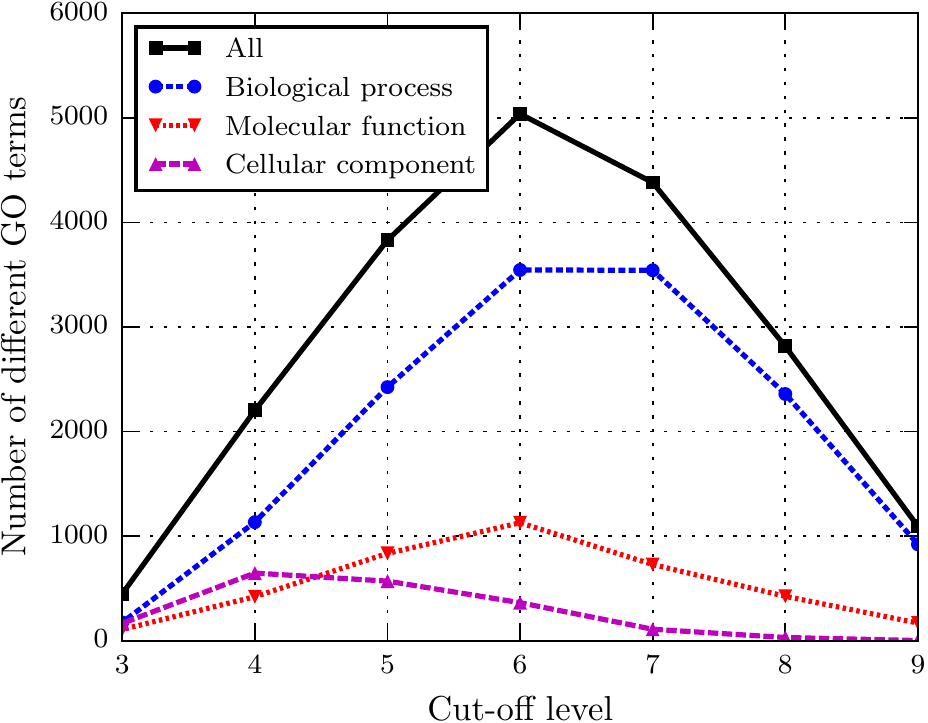}
  \caption{Number of different GO terms for different cut-off levels} 
\label{fig:nb_go_level}
\end{center}
\end{figure}

\begin{figure}[htb!]
\begin{center} 
\includegraphics[width=10cm, 
keepaspectratio]{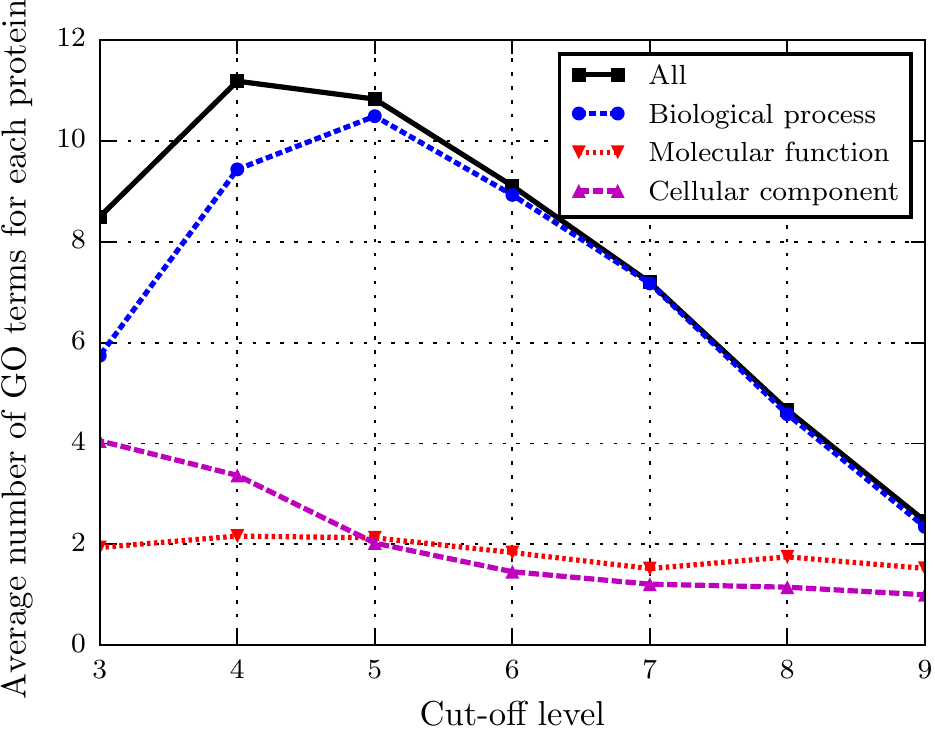}
  \caption{Average number of GO terms for each annotated protein for 
different 
cut-off levels} 
\label{fig:avg_go_per_prot}
\end{center}
\end{figure}

In Figure~\ref{fig:nb_prot_go_level}, the total number of annotated proteins, at
different cut-off levels, are shown. Also, the number of GO terms and the 
average 
number of GO terms for each annotated protein, at different cut-off levels, are 
shown in Figures~\ref{fig:nb_go_level} and \ref{fig:avg_go_per_prot}, 
receptively.

\end{document}